\documentclass[fleqn,usenatbib]{rasti}

\usepackage{newtxtext,newtxmath}

\usepackage[utf8]{inputenc}
\usepackage[T1]{fontenc}

\DeclareRobustCommand{\VAN}[3]{#2}
\let\VANthebibliography\thebibliography
\def\thebibliography{\DeclareRobustCommand{\VAN}[3]{##3}\VANthebibliography}
\newcommand{\sigpsi}{\sigma_\psi}
\newcommand{\chisq}{\chi^2}

\usepackage{graphicx}
\usepackage{amsmath}

\title[An AI-assisted analysis pipeline for IXPE]
{A Pipeline for the AI-Assisted Analysis of Imaging X-ray Polarimetry Explorer Observations of Black Hole X-ray Binaries 
and Limits on Their Polarization Variability}

\author[Domine et al.]{
B.\ Domine$^{1}$, 
B.\ Liu$^{1}$, 
H.\ Krawczynski$^{1,2}$ \thanks{Corresponding author: krawcz@wustl.edu}, 
M.\,Kundu$^{1}$, 
Shravan Vengalil Menon$^{1}$, 
Fang Zhou$^{1}$ \thanks{Corresponding author: f.zhou1@wustl.edu}
\\
$^{1}$ Washington University in St. Louis, Physics Department, 1 Brookings Dr., CB 1105, St. Louis, MO 63105, USA\\
$^{2}$ Washington University in St. Louis, McDonnell Center for the Space Sciences and Center for Quantum Leaps,\\
1 Brookings Dr., CB 1105, St. Louis, MO 63105, USA
}

\pubyear{\the\year{}}

\begin{document}
\label{firstpage}
\pagerange{\pageref{firstpage}--\pageref{lastpage}}
\maketitle

\begin{abstract}
The {\it Imaging X-ray Polarimetry Explorer (IXPE)} has opened a new window into the high-energy Universe, with the capability to measure the arrival time, arrival direction, energy, and, for the first time with the required sensitivity, the polarization of X-rays from black holes, neutron stars, and other cosmic sources of X-rays. In this paper, we discuss the use of Artificial Intelligence (AI) agents for the analysis of {\it IXPE} data. The pipeline combines an immutable code base with evolving knowledge databases. The generation of analysis projects makes use of a literature data base to judge the novelty of the proposed analysis projects. We emphasize that the analysis requires two types of expert knowledge: detailed
instructions about how to reduce the X-ray data, and
expert knowledge about how to interpret the results.
We employ several levels of auditors that scrutinize the analysis plan, the analysis scripts, and the analysis results for data and method adequacy. The pipeline includes a staggered approach to interpret the analysis results. 
We present exemplary results from the pipeline applied to {\it IXPE} observations of Black Hole X-ray Binaries (BHXRBs) including comparisons of results from the analysis pipeline with published results and a systematic search for polarization degree and polarization angle variability in long {\it IXPE} observations. We critically discuss the performance of the pipeline and future avenues for improving it. 
\end{abstract}


\section{Introduction}
\label{s:intro}
Nobel laureate Riccardo Giacconi and coworkers started the current practice of opening the access to the data from space missions to the scientific community.
Referring to the data from the {\it Einstein} X-ray observatory he stated that ``We are convinced that participation by a broad segment of the astronomical community in the utilization of this facility will substantially enhance the scientific benefits of this mission.''
\citep{1979ApJ...230..540G}.
This pioneering move became the common practice with NASA setting up wavelength-specific centers, X-ray at CfA/SAO, infrared at Caltech/IPAC, UV/optical at STScI, followed by the HEASARC in November 1990 (the high-energy archive that serves {\it IXPE}), plus NED and NASA's Astrophysics Data System (ADS) in the early 1990s. The practice was reinforced by the 2013 Office of Science and Technology Policy ``Holdren memorandum'' on public access to federally funded research \citep{holdren2013memorandum}, and more recently by NASA's Transform to Open Science (TOPS) initiative and the 2023 ``Year of Open Science'' 
\citep{2023DPS....5511703B}.
Other agencies followed NASA's laudable example and instituted similar practices.

Although the satellite data are now widely available, publishing state-of-the-art results still requires a steep learning curve with first-time users needing to learn details of the mission as well as the history of how it changed and evolved during its lifetime. It seems to be desirable to enlist the power of artificial intelligence (AI) in general, and Large Language Models (LLMs) \citep{10433480,NIPS2017_3f5ee243} in particular to help scientists with the task of reducing the data. Recent articles reviewing the use of AI and deep learning in astrophysics can be found in \citet{2026ARA&A..64...19T}, and articles about agentic AI (using LLM agents to analyze data) or autonomous AI (relying exclusively on LLM agents) can be found in \citet{plaat2025agentic,li2026agentic}.
Agentic AI and autonomous AI promise to accelerate the pace of scientific discoveries in various ways:\\[0.2ex]  
(a) LLMs help scientists to scrutinize the literature archive and to evaluate the novelty of analysis projects.\\[0.2ex] 
(b) LLMs lower the barrier of researchers engaging in the analysis of the data from satellites \citep[see][for a related discussion]{wang2023scientific}.\\[0.2ex]
(c) AI assisted research makes it possible to follow up on ideas more quickly, reducing the number of analysis projects that are discarded owing to limited personpower \citep[see][for a related discussion]{gottweis2025aicoscientist}.\\[0.2ex]
(d)  Whereas the expertise of scientists is oftentimes limited to a single field, LLMs are knowledgeable in many different fields, and can thus facilitate the application of statistical methods developed in one field in other fields.\\[0.2ex]
(e) LLMs can come up with original research project ideas in various ways, including following up on ideas mentioned in the literature and transfer of analyses developed in different contexts, and following up on obvious leads that were not implemented by scientists owing to subjective bias \citep[see also][]{si2024llmsgeneratenovelresearch}.\\[0.2ex]
(f) The analysis of data from space missions is somewhat mechanistic and can be scripted. However, the analysis oftentimes requires considerable amounts of attention as data formats change during the lifetime of a mission, and detector problems occur and are addressed with somewhat ad-hoc methods and evolving documentation. Combing through the data and the manuals for how to address specific data issues
is a task well suited for automation.\\[0.2ex]
(g) AI can be used to cross check and validate analyses performed by scientists \citep[e.g.,][]{2024arXiv241200431L}.\\[0.4ex]  
However, LLM driven agents have limitations which make it questionable if they can gainfully be employed in the scientific enterprise.
Shortcomings include the following. The analysis of scientific data requires more rigour and stricter adherence to established protocols than many of the applications for which  LLMs and agents have been optimized. Agents may replace actual data with simulated data, or use shortcuts and simplifications (e.g., replacing actual data with proxies derived from actual data, but lacking the detailed information) that void the purpose of the analysis. LLMs hallucinate, i.e., invent context information such as citations that do not exist 
\citep[][]{Huang_2025}. When using gates to enforce certain data analysis guidelines they may devise methods to circumvent those gates. LLMs are not deterministic and running the same analysis pipeline several times may lead to different results. LLMs are constantly updated and improved; thus re-running an analysis at a later stage may lead to different analysis choices. 

More fundamentally, agentic research will impact the profession and the nature of the work of the scientist. Several issues need to be addressed. Without humans knowledgeable about the details of the research of autonomous research pipelines, 
the benefit of the research for humans and for scientific progress is not assured. Furthermore, human validation is at this point crucial in discriminating valuable research results from false or irrelevant research \citep[see also][]{chen2025scienceagentbenchrigorousassessmentlanguage}. However, scrutinizing AI output is a major time investment, and highly skilled scientists will not stay engaged in research if their role is diminished to vetting the output from agents. With LLMs doing the actual analyses, there is the danger of scientists losing control over the quality and reliability of the results. Researchers can be prone to illusions of understanding the results, even though doing the actual work is oftentimes a prerequisite for a true understanding of the data \citep[see][for a related discussion]{messeri2024artificial}. Agents will make it more difficult to educate the next generations of scientists. With agents not attributing the results to the original authors, and authors being ignorant about the provenance of the work, agents endanger some of the procedural cornerstones that have made science a successful enterprise.

Against the backdrop of these fundamental questions and concerns \citep[see also][]{vannoorden2023ai}, we report here on our work with an agentic system that we developed to analyze data from the {\it Imaging X-ray Polarimetry Explorer (IXPE)} \citep{2022JATIS...8b6002W}. 
The code described here is open source and we encourage interested readers to experiment with it and to contribute to it. The code used for this paper has been version-tagged.
This paper focuses on the analysis of {\it IXPE} data. 
The data reduction library is at present {\it IXPE}-specific. 
Its extension to NICER \citep{NICER}, the {\it Neil Gehrels Swift Observatory} X-ray Telescope (XRT) \citep{XRT}, and {\it NuSTAR} \citep{NUSTAR} is in preparation. The pipeline is the first one developed for the analysis of {\it IXPE} data.  
 More general agentic AI systems include the Denario project \citep{2025arXiv251026887V}, the AI Scientist \citep{Lu2024,Yamada2025}, and the  Multi-Agent System for Cosmological Parameter Analysis
\citep[][]{2024arXiv241200431L}.
       
The rest of the paper is organized as follows. 
We describe the design and implementation of the analysis pipeline in Sect.\,\ref{s:design}.
Section \ref{s:results} shows exemplary results from the pipeline. We conclude with a discussion of the results and a description of avenues for future work in Sect.\,\ref{s:discussion}.
\section{Pipeline Design Principles}
\label{s:design}
\begin{figure*}
    \centering
    \includegraphics[width=0.7\linewidth]{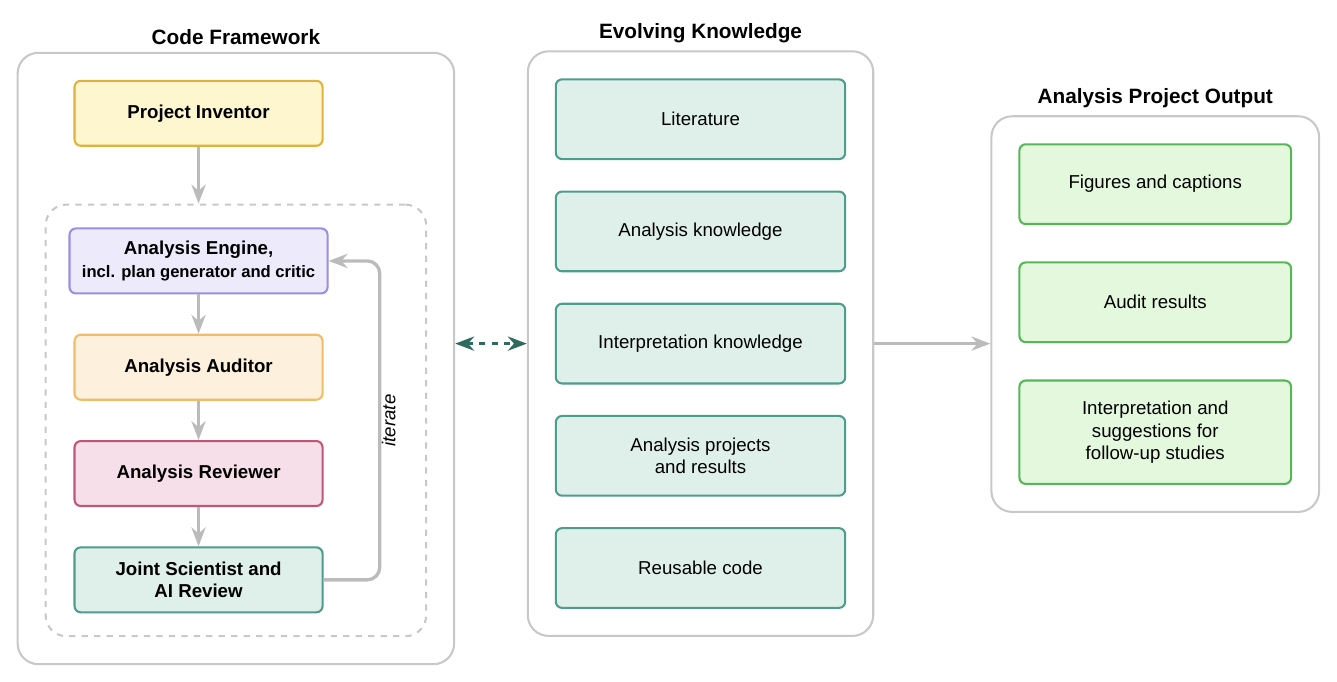}
    \caption{The agentic analysis pipeline uses an immutable code framework (left), an evolving knowledge database (center), and the output for individual analysis projects (right).}
    \label{f:design}
\end{figure*}

The early and exemplary open-data policies of the major space agencies make the astronomical observations with satellite-borne telescopes an excellent testbed for automatic analysis pipelines. The top level design of the analysis pipeline is shown in Fig.\,\ref{f:design}. 
Although the pipeline can run autonomously, actual analyses benefit from the input from scientists, e.g., 
by contributing analysis project ideas, by selecting ideas from the analysis project generator for execution, and by reviewing and adding to the critique of the analysis results, and the formulation of follow-up studies and queries. The pipeline uses four main components (Fig.\,\ref{f:design}): (i) the analysis project generator, (ii) the main analysis engine, (iii) the data and method auditor, and (iv) the interpretation and follow-up engine.  We run (i) once, and then iterate (ii)-(iv) until firm conclusions have been reached.  

We use an immutable code base that uses and refines the knowledge databases. The analysis pipeline runs on a Personal Computer (PC) as {\tt Python} scripts that call the LLM via Application Programming Interfaces (APIs). Given the modest size of the relevant {\it IXPE} black hole and neutron star data archive ($\sim 1$\,TB), we download the entire data set, avoiding the housekeeping involved in fetching the data on an as-needed basis. The actual data analysis is then performed on the scientist's PC via {\tt Python} scripts that the LLM writes. The {\tt Python} scripts are versioned and permanently stored, assuring the reproducibility of the analyses.  

The pipeline uses several types of {\it a priori} information: papers with analysis prescriptions, 
analysis guides, and the run-time help provided 
by the X-ray analysis software {\tt FTOOLS} \citep{1995ASPC...77..367B}, {\tt ixpeobssim} \citep{2022SoftX..1901194B}, and {\tt XSPEC} \citep{1996ASPC..101...17A}.  
It learns through {\bf evolving knowledge databases}: (i) a literature database, (ii) an evolving collection of analysis recipes and data analysis rules (derived from the analysis papers and guides, as well as from previous analysis runs), 
(iii) a {\tt Python} library of scientist-vetted reduction functions ({\tt reduction/red\_ixpe.py} and companions: 
background rejection, source finding, response files, polarization fits), and (iv) summaries of prior results, one per source class ({\tt ixpe\_summaries/}), regenerated from the project results and injected into the project inventor's context.

The division into immutable code and evolving knowledge allows the scientists to keep control of the evolution of the analysis pipeline. When the pipeline makes a major discovery, the scientists need to inspect the analysis scripts before 
signing off on the publication of the results.   

\subsection{Software Environments}
We code the agents with the {\tt LangChain} \citep{langchain2022} and {\tt LangGraph} \citep{langgraph2024} open-source software, 
which allow us to change the LLMs 
used in different roles without changing the code. 
We use the open-standard Model Context Protocol (MCP; \citealp{mcp2024}) to implement the agents' interaction 
with the analysis computer. The {\tt LangChain}, {\tt LangGraph} and MCP code, and the X-ray analysis software require mutually incompatible environments. 
The X-ray analysis software pins a specific {\tt Python} version and a specific {\tt NumPy} version, as the compiled {\tt FTOOLS}/{\tt XSPEC} binaries are built against them. {\tt LangChain}, {\tt LangGraph} and MCP require different versions. 
We thus use two environments, the {\tt agent} environment for the work of the LLM agents and the server of tools that the LLM agents can use,  and the  {\tt analysis} environment for the analysis work that includes the publicly available X-ray data analysis software packages {\tt FTOOLS}, {\tt ixpeobssim}, and {\tt XSPEC}. 
These two environments have to be installed by the user before running the code.
\subsection{Databases}
\label{ss:database}
The databases use either the Markdown format for readability or the JSON (JavaScript Object Notation) format for indexed access. 
We furthermore implement a Retrieval-Augmented Generation (RAG) pipeline to source literature.

A {\tt Python} script extracts information about the {\it IXPE} data into a JSON inventory ({\tt ixpe\_data\_inventory.json}). The inventory stores the observation identifiers (IDs), the source category, the full directory path, the names of the subdirectories, the names of the event files, the detector units (DUs 1--3) that have data, and the total size in MB.

The knowledge about the analysis procedures is stored in a Markdown file ({\tt ixpe\_methods.md}). The pipeline is instructed to prioritize the rules according to the order in which they are presented. The file is extracted automatically from several sources. It starts with general information, i.e.\ the analysis packages that can be used ({\tt FTOOLS}, {\tt ixpeobssim}, {\tt XSPEC}), the rule to try out all analysis on a  small number of observations before analyzing all the data, and file naming conventions. This is followed by the scientist-vetted analysis recipes learned from previous analyses rounds (maintained in {\tt ixpe\_recipes.json}). At the time of writing we use 22 recipes, 
each with an identifier, the trigger phrases for which it applies, a priority, the problem it prevents, the rule, a procedure, a code example, and pitfalls. The script-writing agents are instructed to follow a recipe whenever one of its trigger phrases matches the task, and not to improvise a simpler stand-in. The file furthermore contains detailed command syntax for key {\tt FTOOLS}, {\tt ixpeobssim}, and {\tt XSPEC} commands, and a summary of analysis preferences. We use the paper by  \citet{2026arXiv260403366D} as the highest-authority prescription, and add best practices extracted from 254 papers from ADS which contain the word `IXPE' in the title or abstract.

The knowledge about how to review and interpret results is stored in {\tt ixpe\_critic\_checks.json}. At the time of writing this paper, we use 16 entries. Each entry names a defect pattern (e.g., ``the numbers quoted in a caption must be readable off the figure'', ``an upper limit that scales with the noise is a sensitivity, not a limit'', ``a $\chi^2$ summed over re-binnings of the same data is not a $\chi^2$''), symptom, the diagnosis, known false positives, the suggested follow-up, and the project in which it was first found. The checks are handed to the critics described in Sect.\,\ref{ss:engine}. Both databases grow from run to run. After each run, a recipe-learning agent and a check-learning agent propose additions, which are written to files in the project's results directory ({\tt ixpe\_recipes\_proposed\_addition\_*.json} and {\tt ixpe\_critic\_checks\_proposed\_*.json}), and the scientist merges the accepted entries by hand. The scientist thus stays in control of what the pipeline learns.

The data analysis projects are summarized in {\tt ixpe\_research\_database.json} which stores for each analysis project an identification token (ID), 
a project title,
a list of the involved data sets, a high-level description including the rationale for the analysis, a very detailed step-by-step analysis prescription, and relevant literature references. 
The key insights from previous analyses are stored in per-source-class summary files ({\tt ixpe\_summaries/ixpe\_*\_summary.txt}), which {\tt project\_summary.py} regenerates from the project results.
\subsection{Immutable Code Basis}
\label{ss:immutable}
The immutable code base includes the 
method extractor that derives the analysis rules, the project inventor that generates the analysis projects, and the analysis engine that performs the analysis, audits the methods, and interprets and criticizes the results. We guarantee that the agents' efforts align by providing the agents with identical context information as far as possible. 
\subsubsection*{The Project Inventor}
\label{ss:inventor}
The project inventor generates its own analysis projects or converts an idea described by one of the scientists into an analysis project. The generation proceeds in a generator--critic two-model loop with $N_{\rm inv}\,=\,4$ iterations \citep[see e.g.,][who use an analogous generate-critique-evolve chain for hypothesis generation]{gottweis2025aicoscientist}. 
We confirmed that the iterations lead to a substantial improvement of the quality of the analysis projects in terms of the specificity of the project descriptions and the adherence to our instructions.
The generator and critic are implemented inside a {\tt Python} loop which calls the generator (implemented with Claude, via models Opus 4.8 or Fable 5) and the critic (implemented with Gemini, via the model {\tt gemini-3.1-pro-preview})  as {\tt LangChain} chains with structured output. 
We chose Claude for the generation of analysis prescriptions to make use of its coding and agentic AI prowess. We chose Gemini for the review to take advantage of its large context window, 
and to get a critique that is independent of the 
generator.  We use the {\tt Pydantic} industry standard for enforcing the format of the LLM output \citep{pydantic} and to easily map the data to 
entries in the JSON research database 
described above. 
  
\begin{figure*}
    \centering
    \includegraphics[width=0.7\linewidth]{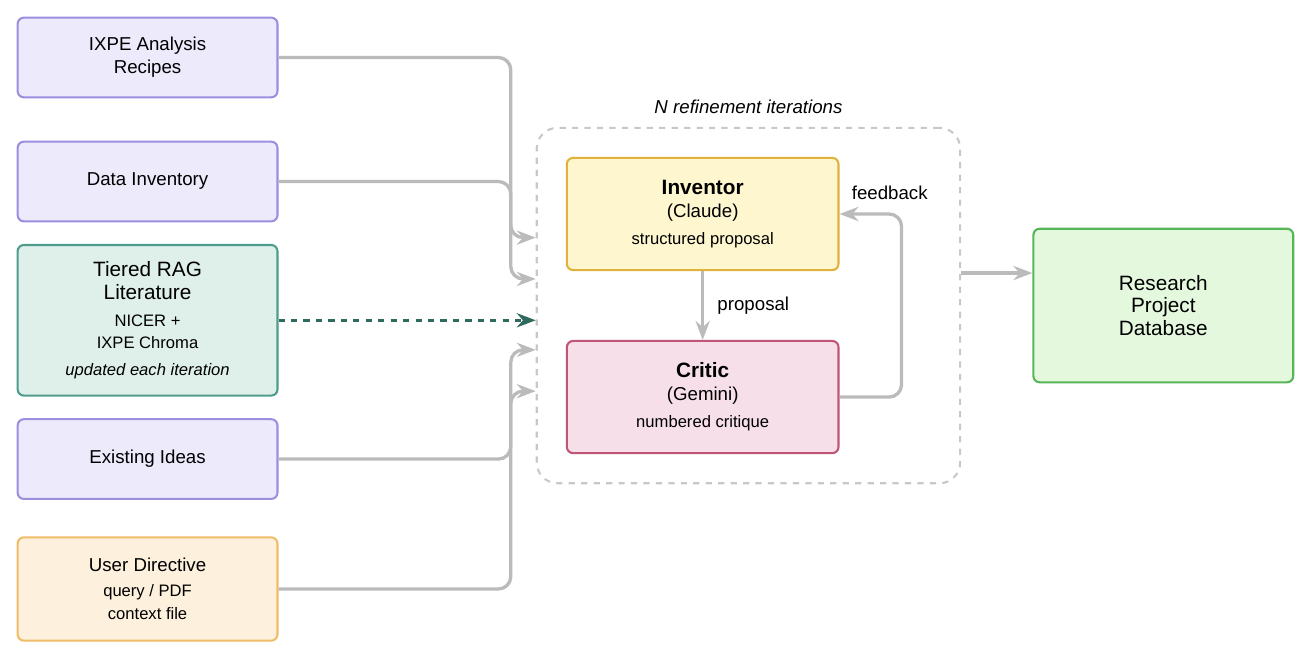}
    \caption{ 
        The generator--critic refinement loop of the project inventor. In each of the $N$ iterations the generator receives optional seed idea and targets, the data inventory, the methods reference, the summaries of prior results, and literature passages retrieved by the RAG module, and writes a structured proposal; the critic reviews the proposal with a larger literature budget, and its review is fed back into the next iteration. After $N$ iterations the final proposal is written to the research project database, from which the analysis engine reads it.} 
    \label{fig:inventorCritic}
\end{figure*}

The context for the project inventor  (Fig.\,\ref{fig:inventorCritic}) includes the execution context that the analysis 
will be performed by an agentic pipeline with 
access to the {\it IXPE} analysis tools ({\tt ixpeobssim}, {\tt HEASoft}/CALDB, {\tt XSPEC}), making use 
of data reduction instructions 
and a data reduction library.  
We provide the project inventor with a tiered 
RAG module \citep{Lewis_2020}, which supplies full passages 
from the most relevant papers and progressively 
lighter summaries for the rest, with a larger 
literature budget for the critic.
The RAG retrieval includes information about the 
original research goal and the current analysis project draft, and the retrieved papers change as the 
proposal evolves across iterations. The project inventor receives a hierarchy of principles that it is asked to respect. The top priority is that the analysis has to be based on actual data and not invented data. The only exception to this rule is simulations of the null hypothesis to ascertain the statistical significance of certain findings.  We instruct the generator and critic to focus on statistical analyses of {\it IXPE} data deriving temporal, spectral, and polarization properties, and to avoid new theoretical studies and/or running of external theoretical model codes. These constraints limit the analysis space, and lead to readily executable projects. We add several options to the project inventor:
the scientist can add the seed idea around which an analysis project is developed, they can add reference literature (``perform an analysis similar to the one described in the included paper''), they can refine 
the project idea and plan interactively, and they can instruct the inventor to import statistical methods from other scientific fields.
\subsection{Downselect of analysis projects}
\label{ss:downselect}
We select the most promising analysis projects before submitting them for execution with the objective to optimize the use of expensive API resources and to optimize the use of the time of the scientists who review the output of the analysis pipeline.
We downselect the project proposals using a three-step ranking algorithm. In the first step, the proposals are filtered through a validity gate. This gate discards 
projects which fail one or more of the following criteria: the analysis project does not include the development of new theoretical models, it includes an explicit analysis prescription, it uses only the standardized toolchain, it only uses the existing data, justifying the source selection, and it accurately references existing literature, e.g., via verifiable links to ADS and the arXiv preprint server.
In the second step, the analysis project plans receive
a score in five categories: publishability, strength of physical constraints, novelty of the proposed work relative to the state of the art and to other prior proposals generated by our pipeline, feasibility and executability of the proposed work (using standard computing resources), and alignment with the {\it IXPE} toolchain. The total score is the weighted score from the five categories with the first three categories being weighted twice as much as the latter two.  
In the third step, the top-ranked $k$ analysis project proposals (we use $k = 12$) are sorted. We use the {\it merge sort} algorithm \citep{knuth1998art} with binary comparisons based on the same five criteria as for the ranking.  Step 3 differs from Step 2 by ranking the two competing proposals in the context of each other rather than on their own. Each comparison is performed twice in reversed order to eliminate position bias. We resolve ties with an independent model, and if this fails as well, with the rubric score.  We use Claude (Fable 5) for steps one and two, and Gemini (Gemini 3.1 Flash) and Claude (Fable 5) for step three.
The non-deterministic nature of an LLM-based comparison method may compromise the integrity of the sort if it fails to uphold the transitive property. 
We tested the transitivity of the comparison method on a random sample of triplets of analysis project plans, and found an astonishingly low intransitivity rate of 4\%.

\subsection{The Analysis Engine}
\label{ss:engine}
The analysis engine converts the analysis project descriptions into executable analysis prescriptions, runs the actual analysis, applies the validation gates, interprets the results in several steps, and makes follow-up suggestions.

It receives the project specification generated by the project inventor including the research objective, the selected targets and observations, the planned analysis steps, relevant literature references, and the current project status. After the generation of a workflow by {\tt claude-opus-4-8}, the critic ({\tt gemini-3.1-pro-preview}) reviews the workflow for adherence to the requested guidelines and hands the workflow and the critique back to the generator. We use three generator--critic iterations to convert the analysis project in this way into an executable data analysis workflow, and two further iterations on the first analysis script. 

With the workflow specified, the analysis engine prompts a {\tt LangChain} ReAct (Reasoning and Acting) agent \citep{yao2023reactsynergizingreasoningacting} ({\tt claude-opus-4-8} for the runs reported here) to perform the analysis. This agent generates {\tt Python} analysis scripts and runs them in the {\it IXPE} analysis environment through the MCP tool server. 
The tool server limits the access of the agent to approved data paths and analysis tools, with the caveat that the agent can write and run its own {\tt Python} scripts which have no constraints on what files they can access. A safe method for keeping private files out of the reach of the agent is to run the analysis pipeline from a dedicated user account with file access 
limited to the analysis directories. 
The MCP server allows the agent to launch and monitor analysis jobs and their CPU and memory usage, to  record logs and exit status, and to collect generated outputs such as figures, tables, provenance records, and summary files.  We prompt the agent to produce result tables and figures with captions, a summary of the research findings, and a provenance file specifying the analysis scripts, input data, and tool calls that produced it. 
The agent is implemented with {\tt LangGraph} and runs asynchronously with all the processes it launches, enabling it to monitor and control these processes.
Without being prompted to do so, the agent reports on the progress of processes, and displays its rationale about the reasonableness of the deployed resources, including order of magnitude estimates of the remaining time to completion. The agent can furthermore find files which are stored at unexpected locations and debug commands that do not work as expected. We implemented tools to minimize unproductive cycles through a loop guard and tool-call timeouts.

\subsubsection{The Analysis Auditor}
In early versions of the analysis engine, the agent tended to replace actual data with simulated data as soon as some problem would show up. Sometimes, it would silently perform a simplified analysis that rendered the analysis useless. An example is an orbital-phase-resolved analysis of the polarization degrees and angles in which the agent took the shortcut to estimate the values in individual phase bins from the results for the entire data set by simple exposure time scaling. The approach would remove any phase dependence present in the data.   
We succeeded in largely eliminating these behaviors by (i) including requests to avoid simulated data and simplified analysis in the workflow generator and critic prompts, and (ii) adding an audit of the results, data and methods after each analysis pass. The analysis auditor reviews the figure captions and provenance descriptions, scans all the analysis scripts, and 
reports infractions to the scientist.
We hide the auditor code and audit results 
from the agent and do not prompt it to redo 
the analysis so that it does not learn how 
to circumvent the auditor. 
\subsubsection{The Analysis Reviewer}
A hallmark of the scientific process is that the initial analysis recipe is necessarily incomplete: 
the data analysis oftentimes yields unexpected results, and those results are likely the most 
interesting ones. Our analysis is thus an iterative process. An initial critic reviews the results one-by-one and comes up with follow-up queries.
We follow the initial interpretation with an analysis by a deep dive agent which ingests the most important figures, extracts hypotheses, and makes or updates a claims register ({\tt claims.json}) which catalogues the main claims with the files from which the evidence stems and their status (active, revised, or superseded); a superseded status is propagated into the associated captions and summaries.
The deep dive agent furthermore refines and refocuses
the follow-up queries with the aims of 
avoiding the snowballing of increasingly irrelevant 
follow-up analyses and of enabling the extraction of scientifically valuable results --- be it detections 
or non-detections. For example, if a systematic search for polarization variability does not give a positive
detection, the deep dive agent generates a follow-up request to characterize the amplitude of the 
polarization variability that the analysis would have detected with statistical significance.
Following the deep-dive agent, the code stops.
The scientist reviews the follow-up queries from the deep dive agent, decides if a follow-up analysis is warranted and if so, comments, edits, or adds the follow-up queries.
We found it to be important that the scientist is involved in the follow-up questions; otherwise they will likely not be convinced that the results are 
watertight.

We added the capability to discuss the analysis results and to strategize about next steps in an interactive scientist--Claude session. The agent can access all analysis results; the scientist can also upload additional articles and request plotting the results 
in different ways. At the end of the discussion session, the scientist can request to update the follow-up analysis description file. The discussion transcript is saved alongside the project files.
\subsubsection{Token Economy}
\label{sss:token-economy}
The number of analyses that the pipeline can run is limited by the time commitment of the scientist checking on the results, and by the LLM API costs. The pipeline implements several context-engineering practices from an Anthropic post \citep{Anthropic2025context}. 
\begin{itemize}
\item Prompt caching with a moving breakpoint: The pipeline makes use of the cost reduction from caching the input as much as possible, as the API costs for cached input are $0.1$ of the full costs. This is implemented by making the conversation append-only, and by adding a movable cache breakpoint to the last message of a request \citep{Anthropic2024caching,Google2024caching,gim2024promptcachemodularattention}. Long cache lifetimes of between 5\,min and 1\,hour are chosen, exceeding the longest blocking waits on analysis jobs. If the provider rejects the cache request, the pipeline sends the  uncached prompt.
\item The pipeline uses short per-call context windows with the help of a pre-model hook \citep{LangGraph2025hook,Liu_2024}. While the 
complete history is stored locally for later access, 
the code sends only the sections of the context relevant for the current request to the LLM. Outdated process-monitoring output is replaced by short placeholders, and the center part of the conversation is trimmed to the budget. The task description and the most recent parts of the conversations are fully transmitted.
\item The windowing of the chat history can lead to the loss of essential information, leading to unproductive ``forget-and-rediscover'' loops. The pipeline therefore uses a scratchpad of essential information that is injected into the LLM prompts \citep{packer2024memgptllmsoperatingsystems,shinn2023reflexionlanguageagentsverbal}.  
\item We limit the output of the LLM whenever possible, and encourage localized edits over file rewrites.
\item Reviewed plans and scripts are passed by file path and executed in place. Analysis products are written as numbered result files with a one-line index and machine-readable provenance tokens that the agent can reference \citep{Aizawa2025tools}.
\item We match the use of different models to the requirements
\citep{chen2023frugalgptuselargelanguage,ong2025routellmlearningroutellms}. The more expensive models are used for the project and workflow generation, and cheaper models for the critic and the summaries. 
\item A loop guard detects unproductive repetitions.
With all of these measures the cost of one iteration of the analysis is usually of the order of 20--40 USD.
New types of analyses that require the agent to debug a problem by trial and error may cost as much as 100 USD.
\end{itemize}
\section{Example Results}
\label{s:results}
\subsection{Analysis Project Plans}
We used the project inventor to generate 43 analysis project plans for {\it IXPE}'s archival black hole X-ray binary (BHXRB) sample.
They can be categorized as follows: 
\begin{itemize}
    \item {Temporal variability of the polarization properties  (19 projects).} The proposed analysis methods include structure-function, autocorrelation, and wavelet analyses, change-point detection, and searches for polarization angle swings, search for periodicities and quasi-periodic modulations.
    \item {Dependence of polarization properties on energy  (9 projects).} Single-source and source-ensemble studies of polarization properties as a function of energy.  

    \item {Correlation of the polarization properties with the flux and hardness (6 projects).} Correlation of the polarization properties with source flux and/or hardness, including investigating features such as hysteresis behavior, geometry of the hardness--intensity track, and directional lead/lag (i.e., do changes in spectral hardness or flux happen first followed by change in polarization properties such as $q$ or $u$, or the other way around). 

    \item {Multi-epoch/population census (5 projects).} Epoch to epoch variability, population clustering, and the homogeneity of the broadband $(q, u)$ distribution. 

    \item {Orbital phase and absorption-driven modulation (3 projects).} Analyzing the dependence of the polarization signal on the binary orbital phase and/or on absorption dips. 
    \item {Reproduction of previous work (1 project, scientist suggested).} A reproduction of the May 2022 {\it IXPE} polarization measurement of Cyg X-1 \citep{Krawczynski2022}, primarily for validation of the pipeline.
\end{itemize}
\begin{table*}
 \centering
 \caption{Comparison of the results of the unweighted analysis of the
 analysis pipeline and the results presented in
 \citet{Krawczynski2022}. PD: polarization degree; PA: polarization angle; DU: detector unit; MC: Monte Carlo.}
 \label{t:cygx1_unweighted}
 \begin{tabular}{lccc}
  \hline
  Quantity & Pipeline & \citet{Krawczynski2022} & Agreement\\
  \hline
  May PD [\%]& $3.80 \pm 0.21$ & $4.01 \pm 0.20$ & $0.73\sigma$\\
  May PA [$^\circ$]& $-21.0 \pm 1.6$ & $-20.7 \pm 1.4$ & $0.15\sigma$\\
  June PD [\%] & $3.97 \pm 0.34$ & $3.84 \pm 0.31$ & $0.28\sigma$\\
  June PA [$^\circ$] & $-26.1 \pm 2.5$ & $-25.7 \pm 2.3$ & $0.11\sigma$\\
  Combined PD [\%]& $3.83 \pm 0.18$ & $3.95 \pm 0.17$ & $0.49\sigma$\\
  Combined PA [$^\circ$] & $-22.3 \pm 1.4$ & $-22.2 \pm 1.2$ & $0.07\sigma$\\
  Detection significance & $21.1\sigma$ & $\sim$$20\sigma$ & --\\
  PD (2--3 / 3--4 / 4--6 / 6--8\,keV) & 3.07 / 3.22 / 4.00 / 5.60 & 3.5 / 3.5 / 4.7 / 5.8 & 0.8 / 0.5 / 1.2 / 0.2$\sigma$\\
  PA (2--3 / 3--4 / 4--6 / 6--8\,keV) [$^\circ$]& $-17.8$ / $-21.4$ / $-21.9$ / $-23.0$ & $-18$ / $-21$ / $-21$ / $-25$ & all $\leq 0.4\sigma$\\
  PD(E) slope $B$ (per cent keV$^{-1}$) & $0.52 \pm 0.12$ & $0.58 \pm 0.15$ & $0.31\sigma$\\
  Constant-PD rejection & $4.2\sigma$ (MC: 3.3--3.5$\sigma$) & $\sim$$3.4\sigma$ & comparable\\
  PA(E) & constant ($p = 0.91$) & constant & consistent\\
  Per-DU PD (May, DU1/2/3) [\%]& 3.66 / 3.80 / 4.00 & 3.8 / 4.2 / 4.1 & 0.2--0.9$\sigma$\\
  Per-DU PA (May, DU1/2/3) [$^\circ$] & $-18.8$ / $-19.0$ / $-25.5$ & $-20$ / $-18$ / $-25$ & $\leq 0.3\sigma$\\
  XSPEC May PD / PA [\%]/[$^\circ$]& $3.55 \pm 0.15$ / $-20.4 \pm 1.2$ & $3.9 \pm 0.2$ / $-21 \pm 1$ & 1.4$\sigma$ / 0.4$\sigma$\\
  Weighted May PD (main run, for reference) [\%]& $3.47 \pm 0.19$ & $4.01 \pm 0.20$ & $1.95\sigma$\\
  \hline
 \end{tabular}
\end{table*}
\begin{figure}
\centering
 \includegraphics[width=0.9\linewidth]{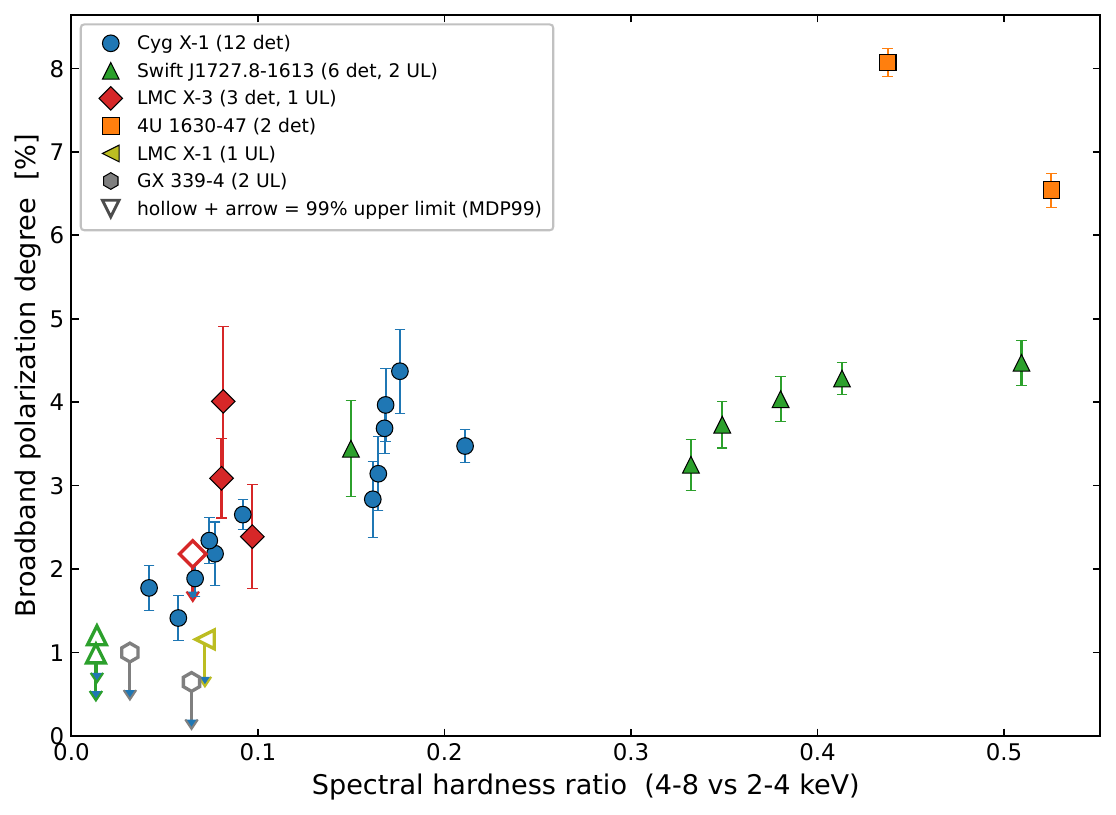}
\caption{Pipeline results for the 2--8~keV polarization degree versus 4--8~keV/2--4~keV hardness ratio correlation. For the sources Cyg X-1 and Swift J1727.8-1613, it shows a positive correlation. The two 4U\,1630$-$47 observations show the opposite trend.}
\label{f:r36}
\end{figure}
\begin{figure*}
    \centering
    \includegraphics[width=0.45\linewidth]{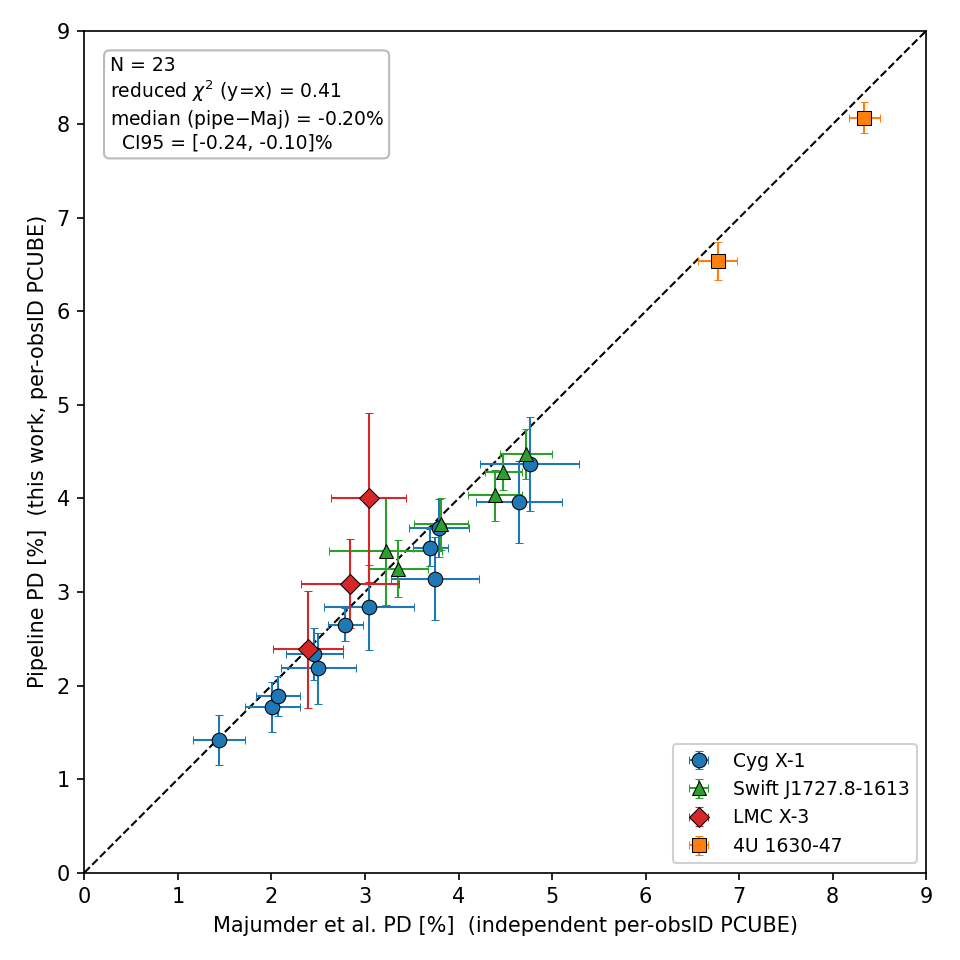}
    \hspace*{0.25cm} \includegraphics[width=0.45\linewidth]{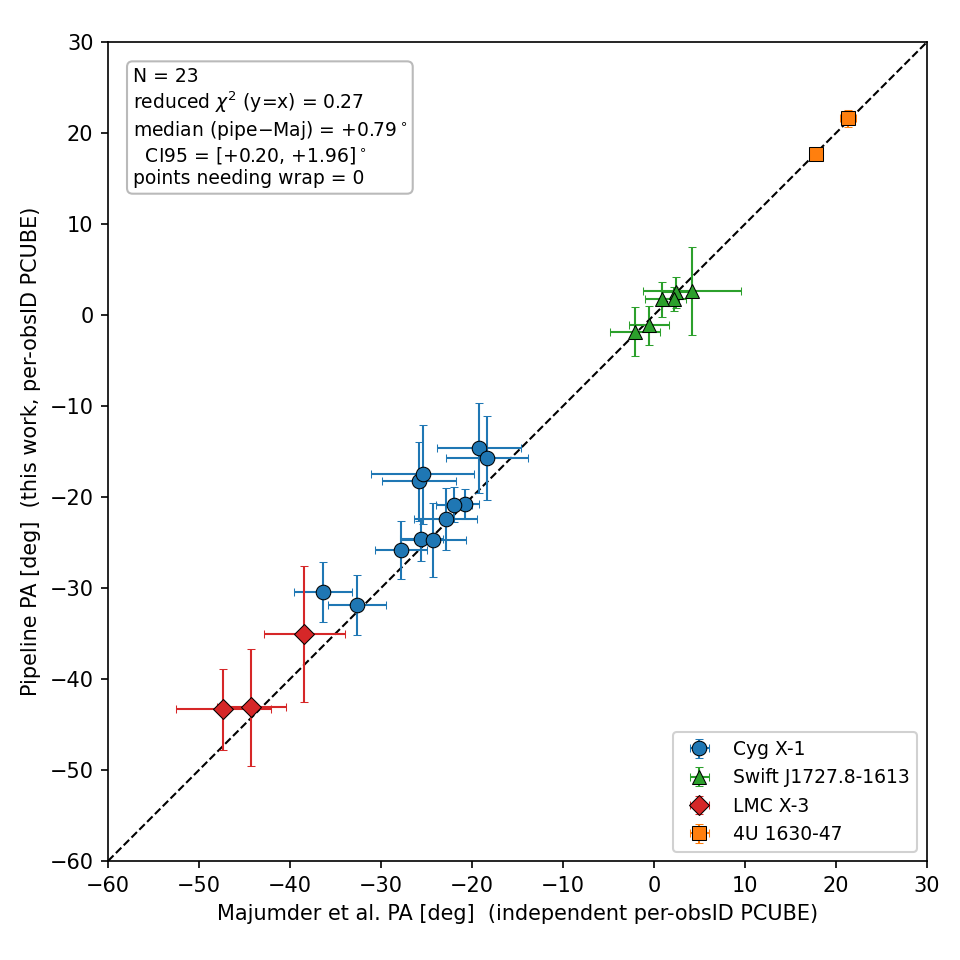}
\caption{Comparison of polarization degrees and angles from the pipeline with the results of \citet{2026MNRAS.545f1933M} (their Table~5; 23 observations with pipeline detections).}
    \label{f:comp}
\end{figure*}
The 10 top-scoring proposals are summarized in Table \ref{t:top10_proposals}.  
\begin{table*}
\centering
\caption{The 10 top-ranked BHXRB analysis proposals. 
The LLM assigned scores P1--P3 are
double weighted, and S1--S2 are single weighted.}
\label{t:top10_proposals}
\small
\renewcommand{\arraystretch}{1.15}
\begin{tabular}{@{}c p{3.2cm} p{7cm}c c c c c c@{}}
\hline
Rank & Proposal title & Brief summary & P1 & P2 & P3 & S1 & S2 & Total \\
\hline
1  & Absorption-driven polarization changes across X-ray dips& Compare dip \& persistent $(q, u)$ in obscured systems& 4 & 4 & 4 & 3 & 5 & 4.000 \\
2  & Energy-dependent $(q, u)$ vector coherence \& rotation& Map energy-resolved Stokes vector arcs across spectral states& 4 & 4 & 4 & 3 & 5 & 4.000 \\
3  &  Orbital phase dependent polarization modulation& Search for phase-locked $(q, u)$ modulation in high mass BH binaries & 4 & 4 & 4 & 2 & 5 & 3.875 \\
4  & Energy-dependent Stokes vector rotation census& Test radial versus rotating $q(E)$, $u(E)$ tracks& 4 & 3 & 4 & 3 & 5 & 3.750 \\
5  & State-dependent PA rotation census& Measure PA-energy rotation \& compare it across states& 4 & 3 & 4 & 3 & 5 & 3.750 \\
6  & Energy-dependence of polarization for different fluxes & Test whether the dependence of Stokes parameters tracks flux \& hardness& 4 & 3 & 4 & 3 & 5 & 3.750 \\
7  & Low frequency periodic polarization modulation& Search for periodic or quasi-periodic $(q, u)$ variability& 3 & 4 & 4 & 2 & 5 & 3.625 \\
8  & Atlas of energy-resolved Stokes trends& Build a cross-source atlas of $q(E)$, $u(E)$ energy gradients& 4 & 2 & 4 & 3 & 5 & 3.500 \\
9  & Orbital phase-resolved polarization via GTI binning& Phase-bin Cyg X-1 or Cyg X-3 events \& test $(Q, U)$ modulation& 3 & 4 & 4 & 2 & 4 & 3.500 \\
10 & Polarization slope versus spectral state study& Fit the energy-dependence of PD \& compare it with disk fraction& 3 & 3 & 4 & 2 & 5 & 3.375 \\
\hline
\end{tabular}

\vspace{0.15cm}
\raggedright
\footnotesize
P1: publishability; P2: physics constraint; P3: novelty;
S1: feasibility; S2: executability.
\end{table*}

\subsection{Pipeline Validation}

In this section we validate the analysis pipeline against the published {\it IXPE} literature on BHXRBs. We tasked the pipeline to re-analyze the Cyg X-1 results published by \citet{Krawczynski2022}, first with and then without (as in the original analysis) event weights.
Most results with weights ({\tt W\_MOM} weights) and without weights are in excellent agreement with each other and with the results of the original paper (Table \ref{t:cygx1_unweighted}). 
The exception is that the inclusion of event weights 
reduces the polarization degree significantly from
($3.80\pm0.21$)\% (this work, no weights) or $(4.01\pm 0.20)$\% (published, no weights) to 
$(3.47\pm 0.19)$\% (this work, with weights).
We reproduced the discrepancy with a manual analysis. 
The results in the next section show that the bias of unweighted versus weighted results is systematic for bright sources.  

One of the first analysis projects that the analysis pipeline generated was a ``Broadband Polarization-Hardness Correlation Survey Across Black Hole X-ray Binaries''. The project included analyzing the observations of the persistent sources Cyg X-1, LMC X-1 and LMC X-3, as well as of the transient sources Swift~J1727.8$-$1613, 4U~1630$-$47 and GX~339$-$4.
The pipeline derived the polarization degrees and hardness ratios, as shown in Fig.\,\ref{f:r36}, using, as most {\it IXPE} analyses do, the newer weighted analysis. This early project used a $3.0'$--$4.0'$ background annulus and subtracted the area-scaled background for all sources, including the bright ones (the default of the reduction library at the time), whereas the analysis of Sect.~\ref{s:res} subtracts the background only for faint sources. The whole-observation polarization degrees of the same 29 observations obtained in Sect.~\ref{s:res} agree with the survey values to better than 0.03\% (absolute), with statistical errors 2--14\% smaller. 
The analysis reviewer diagnosed that the apparent correlation may be an artifact of combining the results from many different sources, and recommended
a follow-up study for the sources with significant detections over a range of hardness ratios, Cyg X-1 and Swift~J1727.8$-$1613. The subsequent follow-up analysis did reveal the well-known correlation (harder energy spectrum - higher polarization degree) for the two sources \citep{2024ApJ...969L..30S,2025A&A...701A.115K}.
Figure \ref{f:comp} compares the pipeline results for the polarization degree and direction with the results published by \citet{2026MNRAS.545f1933M}.
We see a slight but systematic shift of the polarization degrees for the brightest and most polarized sources Cyg X-1 and 4U~1630$-$47. For these two sources, the systematic discrepancy becomes much smaller than the statistical errors when we ask the pipeline to repeat the analysis without event weights. In that case, the results for Cyg X-1 and 4U~1630$-$47 agree perfectly, but those of the weaker sources disagree. 
We believe \citet{2026MNRAS.545f1933M} used the unweighted analysis for Cyg X-1 and 
4U~1630$-$47 and the weighted one for the other sources (the authors do not comment in their paper). 

\subsection{Polarization Variability}
\label{s:res}
Nineteen of the 43 projects scrutinize the time variability of the polarization measured with {\it IXPE}.  The second project proposed by the pipeline was to study the polarization angle (PA) and polarization degree (PD) stability during individual {\it IXPE} BHXRB observations.
The {\it IXPE} BHXRB observations typically run over 1--3 days and have 50--150~ks of live exposure. We used an interactive session to prompt the  inventor to generate a more general project including all ten BHXRBs of the sample, and to scrutinize the variability on different time scales. The analysis uses the normalized Stokes parameters $q$ and $u$ ($-1\le q\le1$, $-1\le u\le1$) giving 
the polarization fraction along two axis systems rotated by 
45 degrees relative to each other (positive values: polarization 
parallel to the axes; negative values: perpendicular to the axes).
The project inventor came up with the idea to scrutinize the Stokes parameters in the rotated frame which maximizes $q'$:
\[
q' = q\cos 2\bar\psi + u\sin 2\bar\psi, \qquad
u' = -q\sin 2\bar\psi + u\cos 2\bar\psi .
\]
where $\bar\psi$ is the PA measured for each observation and energy band individually. For each observation we get $(\bar q', \bar u') = (\mathrm{PD}, 0)$. 
Thus, if only the PD changes while the angle is fixed, only $q'$ varies; 
if only the PA changes at fixed PD, only $u'$ varies.
A small angle variation produces $\delta u' \approx 2\,\mathrm{PD}\,\delta\psi$. A limit on the Root Mean Square (RMS) 
of $u'$ thus gives a limit on the RMS of the PA:
\begin{equation}
    \label{e:PA}
    \sigpsi \approx \frac{\mathrm{RMS}(u')}{2\,\mathrm{PD}} \quad (\text{in radians}).
\end{equation}
The source sample and {\it IXPE} data sets are summarized in Table {\ref{tab:sample}}.
\begin{table*}[h]
\centering\footnotesize
\caption{The source sample of the polarization variability analysis project.
``obs-IDs (FS)'' gives the number of ``frame secure'' (FS) observations with the 2--8~keV polarization detected at the $\ge 5\sigma$ level, allowing for the rotation of the Stokes parameters into the frame that maximizes $q$; 
the upper limits of Fig.~\ref{fig:fig1} use frame-independent tests for the ``poorly defined'' observations, whereas the constancy tests of Fig.~\ref{res:logp} and Table~\ref{tab:top10} are made in the rotated frame for all 34 observations (Sect.~\ref{s:res}). $T_{\rm live}$ is the cumulative live integration time of the data that entered the analysis. State labels (from \citealt{2026MNRAS.545f1933M}; ``--'' = not classified there): LHS = low/hard, HIMS/SIMS = hard/soft intermediate, HSS = high/soft, SPL = steep power law.}
\label{tab:sample}
\setlength{\tabcolsep}{4pt}
\begin{tabular}{lrcrrrl}
\hline
Source & $N_{\rm obs}$ & PD (2--8 keV, \%) & obs-IDs (FS) & \multicolumn{2}{c}{$T_{\rm live}$ (ks)} & States \\
 & & & & all & FS & \\
\hline
Cyg X-1 & 12 & 1.4--4.4 & 12 & 788 & 788 & LHS, HSS \\
Swift~J1727.8$-$1613 & 8 & 0.05--4.5 & 6 & 530 & 314 & HIMS, SIMS, LHS, HSS \\
4U~1630$-$47 & 2 & 6.5--8.1 & 2 & 595 & 595 & HSS, SPL \\
LMC X-3 & 4 & 1.9--4.0 & 1 & 990 & 388 & HSS \\
4U 1957+11 & 2 & 1.7--1.8 & 1 & 707 & 560 & HSS, -- \\
GX~339$-$4 & 2 & 0.3--0.4 & 0 & 191 & 0 & SIMS, HSS \\
LMC X-1 & 1 & 1.1 & 0 & 563 & 0 & HSS \\
Swift~J151857.0$-$572147 & 1 & 0.3 & 0 & 96 & 0 & HSS \\
GS~1354$-$64 & 1 & 3.5 & 1 & 148 & 148 & -- \\
IGR~J17091$-$3624 & 1 & 8.3 & 1 & 163 & 163 & LHS \\
\hline
Total & 34 & & 24 & 4771 & 2956 & \\
\hline
\end{tabular}
\end{table*}
\subsubsection{Data Reduction}
The analysis was refined in 13 follow-up iterations (14 pipeline runs). 
The pipeline did most of the work: it wrote and ran all scripts, kept the bookkeeping straight, and proposed most of the follow-up questions. 
Several important corrections, however, came from inspecting the pipeline results. We noticed that the ``99\% upper limits'' of the first three rounds were detection sensitivities (the amplitude that is detected half of the time). The same mistake returned twice in follow-up analyses. 
In the eighth iteration, we scrutinized the scripts and found 10 shortcomings that the critic has missed as it has access to figures and captions but not the {\tt Python code}. A few suspiciously low count rate points in
earlier versions of Figures \ref{f:4u} and \ref{f:lmcx1} drew our attention 
to the fact that the pipeline had divided the counts 
by the nominal window length rather than by time exposure. 
The rise of the PD of 4U~1630$-$47, the most interesting result of the project, was found by looking at a figure from the first round, after the pipeline had discarded it, and made it less significant by inflating the error bars. In total, the 13 question files contained 370 questions, 126 of which the pipeline executed as follow-up analyses. 

The rest of this paragraph and the next section describe the refined analysis. The analysis uses the scientist-supplied reduction library to 
perform the instrumental background rejection (a filter on the 
shape of the event track that removes particle-induced events). 
Source events are selected  from a circle of radius $1'$ 
around the catalogue position; background from an annulus of 
$2.5'$--$4.0'$. The polarization 
is extracted with {\tt ixpeobssim}'s {\tt xpbin} in PCUBE mode 
with \emph{event weighting} ({\tt W\_MOM} weights: each photoelectron track is weighted by how well its direction can be reconstructed),
with the matching weighted response files. For each energy bin, 
the PCUBE gives the Stokes parameters $I, Q, U$ with errors, 
PD, PA, and the minimum detectable polarization MDP$_{99}$.
The data are analyzed in the 2--4 and 4--8~keV energy bands, and the Stokes
parameters are added to obtain the 2--8~keV results.
Following \citet{2026arXiv260403366D}, the background is subtracted for observations with 2--8~keV source rates $\le $1~count~s$^{-1}$~arcmin$^{-2}$ in the source circles of all three detector units. This applied to five observations: IGR~J17091$-$3624 (04250201), 4U~1957$+$11 (04251199), and three of the four LMC~X-3 observations (02006599, 03004899, 03004901). All other observations, including those of the bright sources Cyg~X-1, Swift~J1727, 4U~1630$-$47 and GS~1354$-$64, are not background-subtracted. The area-scaled background amounts to $\le 1.3$\% of the source counts in every observation, so this choice changes the polarization degrees by at most 1\% of their values.
For every observation the pipeline first makes a sky image with the regions drawn, a light curve of the source and background rates in 300~s bins, and a table. 
A 300~s bin is called a flare if the background rate 
exceeds the median by more than 4.45 times the median absolute deviation (MAD; corresponding to $3 \sigma$). 
Flare intervals are removed from the good-time intervals. 
An observation passes if the source is detected at $I/\sigma_I \ge 5$ over the whole observation. 
In total, 74 flare windows of 300\,s were removed, 
38 of them in one observation of 4U~1957$+$11 (obs-ID 02006601). For each observation and band the whole-observation $\bar q,\bar u$, PD, PA and their errors are computed from one PCUBE. The rotation angle $\bar\psi$ of the mean-vector frame is the whole-observation PA of that band. If PD$/\sigma_{\rm PD} \ge 5$ the frame is called frame secure (FS); otherwise, the observation is flagged ``rotation frame poorly defined'', and the polarization
variability is calculated frame independently.
Each observation is binned into 5\,ks windows, and windows with less than 2.5\,ks of window time inside the flare-free intervals are dropped. For every surviving window, PCUBE Stokes parameters $I_i$, $\sigma_{I,i}$, $Q_i$, $U_i$ and their errors are calculated. Time bins are only used if $I_i/\sigma_{I,i} \ge 5$ in the considered energy band.
Longer time scales are probed by summing the $I,Q,U$ 
(errors added in quadrature) results of the contributing 
bins. Bins which include data gaps exceeding $3\times$ the rung length are excluded from the analysis.

For each time series ($q'$, $u'$) a weighted constant fit is made: the best constant is $c = \sum_i w_i x_i / \sum_i w_i$ with $w_i = 1/\sigma_i^2$, and the goodness of fit is
\[
\chisq = \sum_i \frac{(x_i - c)^2}{\sigma_i^2}, \qquad \text{with } N-1 \text{ degrees of freedom}.
\]
If the polarization is constant, $\chisq$ is distributed like a $\chisq$ variable with $N-1$ degrees of freedom (its mean is $N-1$). The {$p$-value} $p = P(\chisq_{N-1} \ge \chisq_{\rm obs})$ is the probability of getting a scatter at least this large by chance. 

\subsubsection{Statistical Methods}
{\bf Trials correction:} The pipeline corrects the $p$-values for trials with the Bonferroni and the Benjamini--Hochberg (BH) methods \citep{benjamini1995controlling}.  
Considering only the tests with 5\,ks time bins, the number of independent trials is 204 (34 observations, 3 energy bands, and two tests each, one for $q'$ and one for $u'$). Counting all tested time scales, the number of trials is 720. Note that the 720 trials are not all independent of each other. 
The census includes the rotated-frame tests of the ten observations whose frame is ``poorly defined''; for these, $q'$ and $u'$ are fixed linear combinations of $q$ and $u$, set by a whole-observation PA of low significance, rather than PD and PA components, and the frame-independent test is the main result.
The Bonferroni method keeps the probability of at least one false alarm among the $m$ tests below $\alpha = 0.05$ by requiring $p \le \alpha/m$ for an individual test,
i.e.\ $p \le 6.9\times10^{-5}$ for $m = 720$, whatever the dependence between the tests. The BH method keeps the expected fraction of false alarms among the declared discoveries (the false discovery rate, FDR) at or below
$\alpha = 0.05$: it sorts the $m$ $p$-values, $p_{(1)} \le p_{(2)} \le \dots \le p_{(m)}$, finds the largest $k$ with $p_{(k)} \le \alpha\, k/m$, and declares the $k$ tests with the smallest $p$-values discoveries. 

{\bf Variability Limits:} 
For each test (observation, energy band, time scale $\tau$) the pipeline fits
a constant to the $N$ values of $q'$ (and separately $u'$) with weights
$\sigma_i^{-2}$, giving $\chi^2_{\rm obs}$ with $N-1$ degrees of freedom. A
test without detected variability is characterized by three quantities, of
which only the third is quoted as an upper limit.
A bin-to-bin jitter of RMS amplitude $a$ raises the per-bin variance to
$\sigma_i^2+a^2$ and thus the expected $\chi^2$. The sensitivity
$a_{\rm sens}$ is the amplitude at which the \emph{expected} $\chi^2$ reaches
$\chi^2_{0.99}(N-1)$; such a jitter would be detected at $p<0.01$ in about
half of all realizations. 
Like the minimum detectable polarization,
$a_{\rm sens}$ depends only on the error bars and the number of bins, not on
the data, and is not an upper limit. 
Injection tests performed by the pipeline 
with white jitter, sinusoids, and linear ramps recover the 50\% detection amplitude within
$\pm10$\% of $a_{\rm sens}$ at every $\tau$.
The 99\% confidence-level (CL) upper limit $a_{99}$ is the amplitude for which a
$\chi^2$ as small as the observed one has a probability of 1\%,
\begin{equation}
P\!\left[\chi^2(a_{99}) \le \chi^2_{\rm obs}\right] = 0.01 ,
\end{equation}
with $\chi^2(a)$ the constant-fit statistic of data simulated with variances
$\sigma_i^2+a^2$ (common random numbers for all trial amplitudes, root found
by bisection). Injecting jitter of $0$--$2\times a_{99}$ into the headline
cells confirms the 99\% coverage. When $\chi^2_{\rm obs}$ fluctuates below
its null expectation, however, $a_{99}$ becomes anomalously tight, and it
vanishes for the 1\% of tests with $\chi^2_{\rm obs}$ below the 1\% quantile
of the null distribution.

Following \citet{cowan2011powerconstrainedlimits},
the pipeline calculates the RMS limit that it can set in the absence of a signal and then quotes the ``power-constrained limit'', defined as the larger of the limit calculated from the actual data set, and the limit in the absence of a signal:
\begin{equation}
a_{\rm PCL} = \max\!\left(a_{99},\; a_{99}^{\rm med}\right),
\end{equation}
where $a_{99}^{\rm med}$ is the limit obtained for $\chi^2_{\rm obs}$ equal
to the median of the null $\chi^2_{N-1}$ distribution. 
About half of all
tests are quoted at $a_{99}^{\rm med}$, and their limits are determined by the error bars and bin count. 
Using $a_{99}^{\rm med}$ rather than 
the $-1\sigma$ expected limit of \citet{cowan2011powerconstrainedlimits} is a
conservative choice. 
All upper limits given in this paper are power-constrained
99\% CL limits.
For sources with several observations, the constant fits are made per observation in its own rotated frame and $\chi^2$ and degrees of freedom are summed. 
Limits on $q'$ are given as a fraction of the polarization degree and limits on $u'$ as an RMS polarization
angle (Equation \ref{e:PA}).

Three sources (GX~339$-$4, LMC~X-1, Swift~J151857) have PDs 
too small for determining a valid PA. For them a rotation-independent test is used: $\chisq = \sum_i [(q_i-\bar q)^2 + (u_i-\bar u)^2]/\sigma_i^2$ with $2(N-1)$ degrees of freedom, and the same limit machinery gives a 99\% CL limit on the per-component jitter RMS in absolute normalized-Stokes units. 

As mentioned above, the analysis agent decided to ``calibrate'' the error bars, i.e., to inflate the errors of each (energy band, time scale) group by a factor $f = \sqrt{\langle\chi^2/{\rm dof}\rangle}$ whenever the group's mean reduced $\chi^2$ exceeded unity, before deriving the upper limits (the $\chi^2$ values and $p$-values of the constancy tests themselves were always computed with the errors as measured). Because $f$ is set by the most variable observations of a group, this removes real polarization variability along with any error mis-estimate: the 80 and 160~ks factors of the 2--8~keV band (1.11 and 1.47) were dominated by the 4U~1630$-$47 variability discussed below. We therefore instructed the pipeline to compute all limits and significances of record with the errors as measured, and to report the effect of the inflation as a note (e.g., the $p_{\rm chance}$(adj.) column of Table \ref{tab:top10}).
\subsubsection{Results}
The token cost for the initial analysis was 13.80\,USD. We subsequently ran the pipeline 13 times on its own follow-up suggestions and a few of our own follow-up suggestions. 
Once, a follow-up run tripped the data-fabrication trigger. The agent had lost the actual results as it only received the last 40 lines of the output from a tool, and those were dominated by unexpected information. 
The agent then made up a result in its written answer to a review question. Whereas the critic would usually have caught that mistake, it prematurely aborted when trying to ingest an oversized figure. 
The auditor agent caught the mistake and aborted the analysis engine. We subsequently removed all results from this iteration, and restarted the pipeline only after implementing a series of important improvements: scripts return all science lines (rather than the last 40 lines), all prose (answers, captions, conclusions) is 
now checked against the provenance sidecars, result files and run logs, and failed critic turns are tagged.
With all 14 runs, the total cost of the analysis project was 361\,USD (380\,USD including the quarantined run), for 2426 model calls; 59\% of the cost went to the analysis agent, 18\% to the grounding gate, 13\% to the critics, and the rest to bookkeeping and integrity audits. The costs were higher than necessary, as we asked the agent to regenerate all key results and captions at the end of each of the 13 analysis turns. We are working on changing the code, i.e., regenerate results only when called with the flag ``final'', which should reduce the costs of an entire project well below 100\,USD.

\begin{figure}
\centering
 \includegraphics[width=0.95\linewidth]{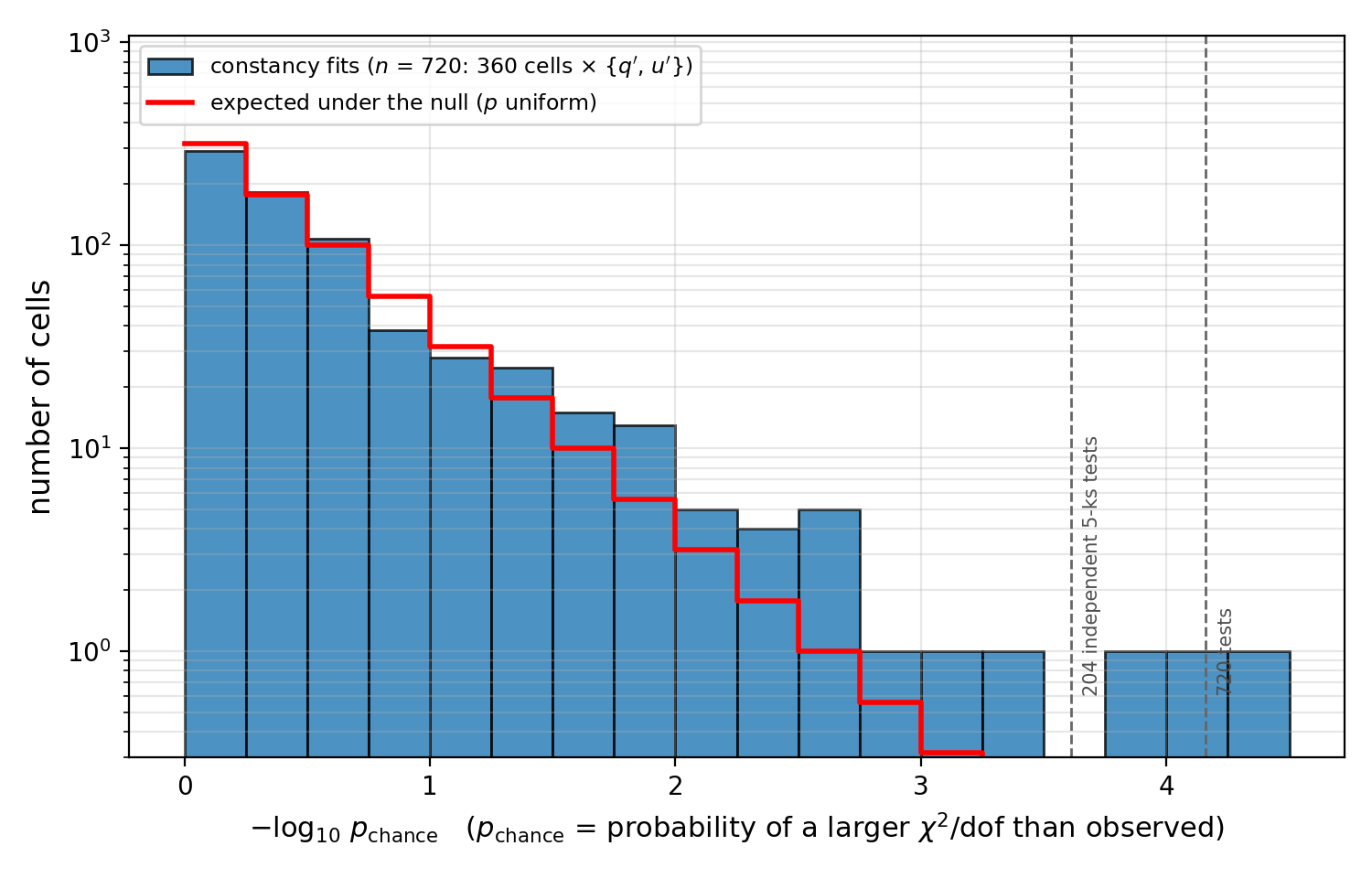}
\caption{Results from the search for polarization variability for all ten sources. The figure shows the distribution of $-\log_{10}p_{\rm chance}$ of all 720 searches ($p_{\rm chance}$ = the probability of a larger $\chisq/{\rm dof}$ than observed), against the expectation for uniform $p$ (red) and the Bonferroni thresholds for 720 tests and for the 204 independent 5~ks tests. The five most significant tests were found in order of decreasing significance for:  
4U~1630--47 01250401 2--8~keV at 80 and 160~ks ($q'$; the two Bonferroni survivors), LMC~X-1 obs-ID 02001901 2--4~keV 40~ks ($q'$), Cyg~X-1 obs-ID 01002901 2--4~keV 10~ks ($q'$), and 4U~1630$-$47 obs-ID 01250401 2--8~keV 40~ks ($q'$). Twenty tests have $p<0.01$ against 7.2 expected, five have $p<0.001$ against 0.7.}
\label{res:logp}
\end{figure}

The first main result of the search, the distribution of the $p$-values (more precisely, the $-\log_{10}p_{\rm chance}$-values) of all searches
is shown in Figure \ref{res:logp}. The pipeline had initially plotted this as $\chi^2 / {\rm dof}$, but we instructed it to show the results as a function of the $p$-values, as the $\chi^2 / {\rm dof}$ values do not map directly to statistical significance. We list the results for the ten most significant fits in Table \ref{tab:top10}.   
\begin{table*}
\centering
\caption{The ten most significant constant-fit tests of the 720, most significant first. $p_{\rm chance}$(nom) is the chance probability of a larger $\chisq/{\rm dof}$ with the errors as measured.
All other columns use the nominal errors as well, except the column $p_{\rm chance}$(adj.) shows the chance probability with the errors multiplied by the factor $f$ from an error calibration. 
The Bonferroni thresholds are $p \le 6.9\times10^{-5}$ over 720 tests and $2.5\times10^{-4}$ over the 204 independent 5~ks tests; ``BH'' marks the Benjamini--Hochberg discoveries at $\alpha=0.05$ over 720.}
\label{tab:top10}
\setlength{\tabcolsep}{2.2pt}
\begin{tabular}{rllcrrcrcccccl}
\hline
 & Source & ObsID & Band & $\tau$ & $N$ & Comp. & $\chisq$/dof & $f$ & $p_{\rm chance}$ & $p_{\rm chance}$ & $-\log_{10}p$ & $-\log_{10}p$ & Detection \\
 & & & (keV) & (ks) & & & & & (nom) & (adj.) & (nom) & (adj.) & \\
\hline
1 & 4U 1630--47 & 01250401 & 2--8 & 80 & 10 & $q'$ & 35.2/9 & 1.107 & $5.6\times10^{-5}$ & $7.3\times10^{-4}$ & 4.26 & 3.14 & Bonf.~720, BH \\
2 & 4U 1630--47 & 01250401 & 2--8 & 160 & 5 & $q'$ & 24.7/4 & 1.471 & $5.9\times10^{-5}$ & $2.2\times10^{-2}$ & 4.23 & 1.65 & Bonf.~720, BH \\
3 & LMC X-1 & 02001901 & 2--4 & 40 & 18 & $q'$ & 46.5/17 & 1.094 & $1.5\times10^{-4}$ & $1.9\times10^{-3}$ & 3.83 & 2.72 & BH \\
4 & Cyg X-1 & 01002901 & 2--4 & 10 & 49 & $q'$ & 87.5/48 & 1.093 & $4.4\times10^{-4}$ & $1.1\times10^{-2}$ & 3.36 & 1.96 & -- \\
5 & 4U 1630--47 & 01250401 & 2--8 & 40 & 20 & $q'$ & 44.1/19 & 1.030 & $9.2\times10^{-4}$ & $2.0\times10^{-3}$ & 3.03 & 2.69 & -- \\
6 & Swift J1727 & 03005801 & 2--4 & 10 & 38 & $q'$ & 69.0/37 & 1.093 & $1.1\times10^{-3}$ & $1.6\times10^{-2}$ & 2.96 & 1.79 & -- \\
7 & 4U 1630--47 & 01250401 & 4--8 & 160 & 5 & $q'$ & 16.7/4 & 1.171 & $2.2\times10^{-3}$ & $1.6\times10^{-2}$ & 2.66 & 1.80 & -- \\
8 & Swift J1727 & 03005801 & 2--4 & 80 & 4 & $u'$ & 14.6/3 & 1.188 & $2.2\times10^{-3}$ & $1.6\times10^{-2}$ & 2.65 & 1.79 & -- \\
9 & LMC X-1 & 02001901 & 2--4 & 160 & 4 & $q'$ & 14.5/3 & 1.346 & $2.3\times10^{-3}$ & $4.6\times10^{-2}$ & 2.64 & 1.34 & -- \\
10 & Cyg X-1 & 01002901 & 2--4 & 5 & 98 & $q'$ & 140.9/97 & 1.060 & $2.4\times10^{-3}$ & $2.8\times10^{-2}$ & 2.62 & 1.56 & -- \\
\hline
\end{tabular}
\end{table*}
Twenty fits have a $p$-value $<$ 0.01 against 7.2 expected under the null hypothesis, and five have $p < 0.001$ against 0.7 expected.

\begin{figure*}
\centering
 \includegraphics[width=0.7\linewidth]
 {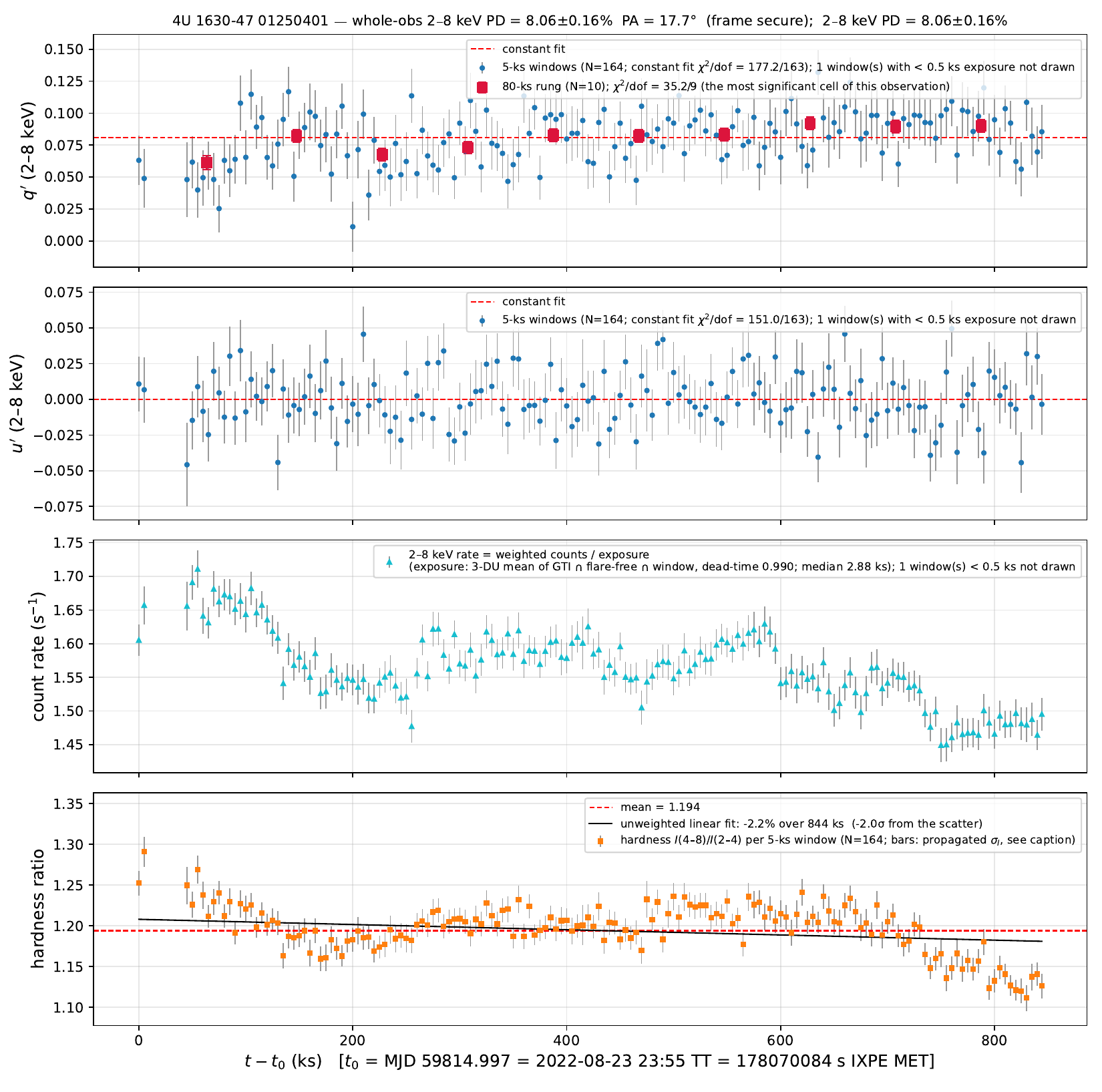}
 \caption{The 4U~1630$-$47 soft-state observation shows significant variability of $q'$ (top panel), but not of $u'$ (second panel from top). $q'$ seems to increase while the weighted counts per second of exposure
 (approximately 11 times lower than the photon count rate), 
 and the weighted 4--8\,keV/2--4\,keV hardness ratios (HRs) decrease.}
\label{f:4u}
\end{figure*}
The 2--8~keV soft-state observation of 4U~1630$-$47 (obs-ID 01250401, Fig.\,\ref{f:4u}) gave the two most significant results for $q'$, measured with 80\,ks and 160\,ks bins, and passed the Bonferroni and BH thresholds. The pipeline produced the
$q'$ and $u'$ histories for all time binnings. 
We used an interactive Claude session to use the 
same data to plot not only $q'$ and $u'$ but also the count rate and hardness ratios, both computed from the {\tt W\_MOM}-weighted intensities of the 5\,ks PCUBEs (the weighted count rate is about an order of magnitude below the photon count rate). 
A weighted straight-line fit to the 164 windows gives a slope of $(3.25 \pm 0.66)\times10^{-5}$ per ks, i.e., a rise of the PD from 6.6\% to 9.4\% over the 844~ks of the observation, at $4.9\sigma$ ($\Delta\chi^2 = 24$ relative to a constant; $3.0$--$4.8\sigma$ with a block bootstrap of the residuals). The same slope is found in each of the three DUs separately ($3.57$, $2.74$ and $3.47\times10^{-5}$ per ks, each $\pm1.15$; $\chi^2 = 0.32$ for 2 degrees of freedom), and $u'$ shows no trend ($-0.1\sigma$): the PD rises while the PA stays constant. We do not see a significant variation of $u'$; we see on average a slight decrease of the count rate (by 8\%), 
and a coherent wander of the 4--8\,keV to 
2--4\,keV HRs. 
$q'$, the count rate and the HRs 
do not show statistically significant 
(Spearman or Pearson) correlations when 
their coherent wander is accounted for.
The time variability of the 
PD in this observation had already been noted by \citet{2024ApJ...964...77R}. 
The results indicate that changes in the emitting plasma or magnetic field geometries can 
affect the PDs, fluxes, and HRs in different ways.

\begin{figure*}
\centering
 \includegraphics[width=0.7\linewidth]
 {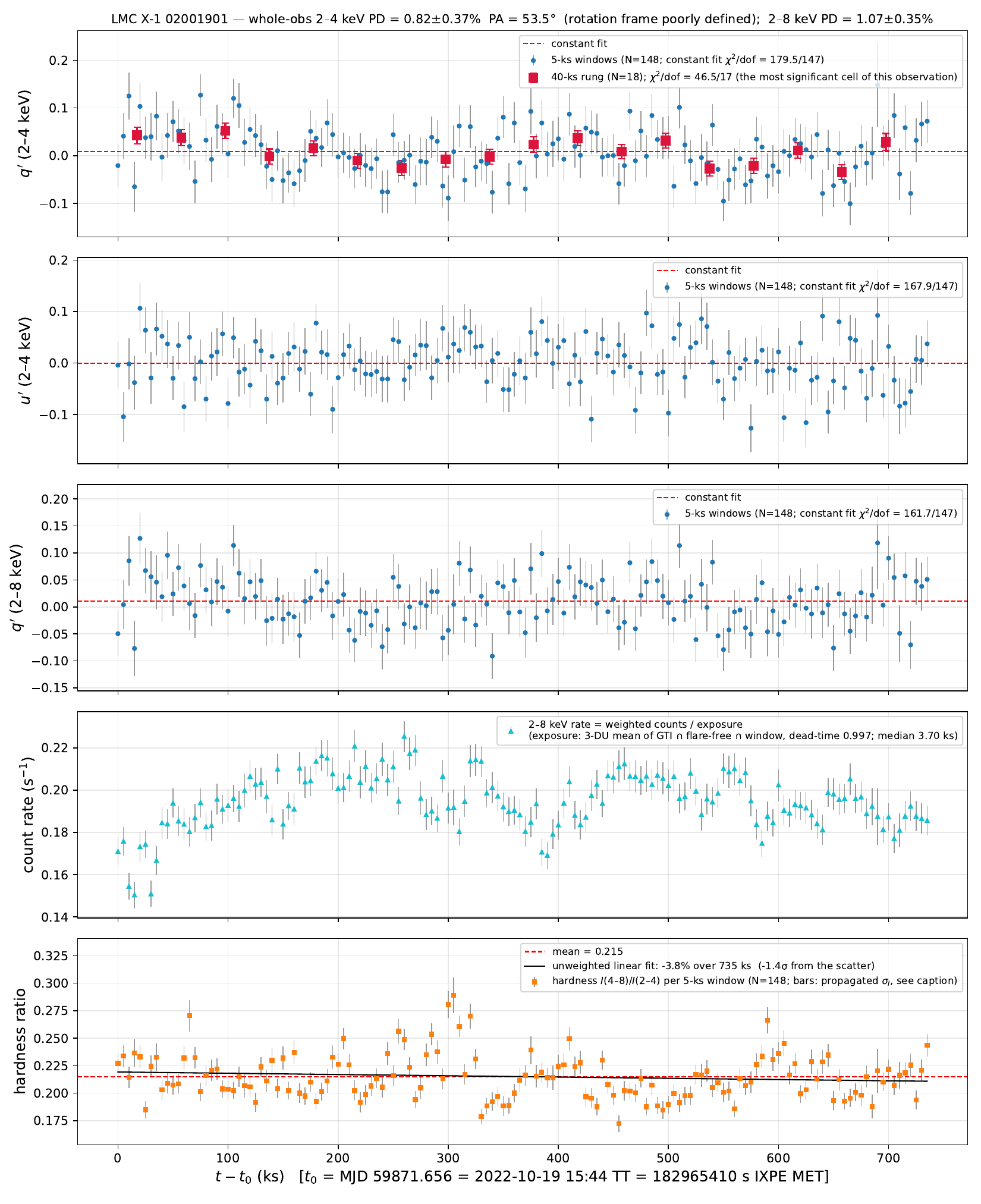}
 \caption{Similar to Figure \ref{f:4u} for LMC X-1, 
 but showing the 2--4\,keV $q'$ and $u'$ curves in the top two panels and the 2--8\,keV $q'$ in the third panel from the top. The $\chi^2$ test with 40\,ks time bins indicates that the 2--4\,keV Stokes parameters are variable. Note that the rotation frame of this observation is poorly defined (2--4\,keV PD/$\sigma_{\rm PD}$ = 2.2), so that $q'$ and $u'$ are fixed linear combinations of $q$ and $u$ rather than well defined PD and PA components. The count rate and hardness are again computed from the {\tt W\_MOM}-weighted intensities. The photon rate in the $1'$ circle: $\approx$3.7~counts~s$^{-1}$ for the three detector units.}
\label{f:lmcx1}
\end{figure*}
Figure \ref{f:lmcx1} shows the results for the next significant observation of LMC X-1. The change of the 2--4\,keV $q'$-values is significant according to the BH criterion, but does not meet the more stringent Bonferroni criterion. The rotation frame of this observation is poorly defined (whole-observation PD/$\sigma_{\rm PD}$ = 2.2 at 2--4\,keV and 3.0 at 2--8\,keV; Table~\ref{tab:sample}), so the $q'$ test is a test of one somewhat randomly fixed linear combination of $q$ and $u$. The frame-independent test of the same 40\,ks rung gives $\chisq/{\rm dof} = 61.6/34$ ($p = 2.6\times10^{-3}$); in the detector frame the variability is carried by $u$ ($\chisq/{\rm dof} = 44.3/17$) rather than $q$ (17.3/17). 
Again here, the $q'$ wander without a clear correlation with the count rate or the hardness ratio.  
We conclude the discussion of the search results by noting (i) the excess of small $p$-values above $-\log_{10}p_{\rm chance}$-values of 1.5, 
indicating a low level of $q'$ variability, and (ii) that 
only one (Swift~J1727, obs-ID 03005801, 80~ks binning) of the 10 most significant fits concerned $u'$ rather than $q'$. We do not show the 
results here as the random wander does not increase the confidence in true $u'$ variability.    

\begin{figure*}
\centering
 \includegraphics[width=0.7\linewidth]
 {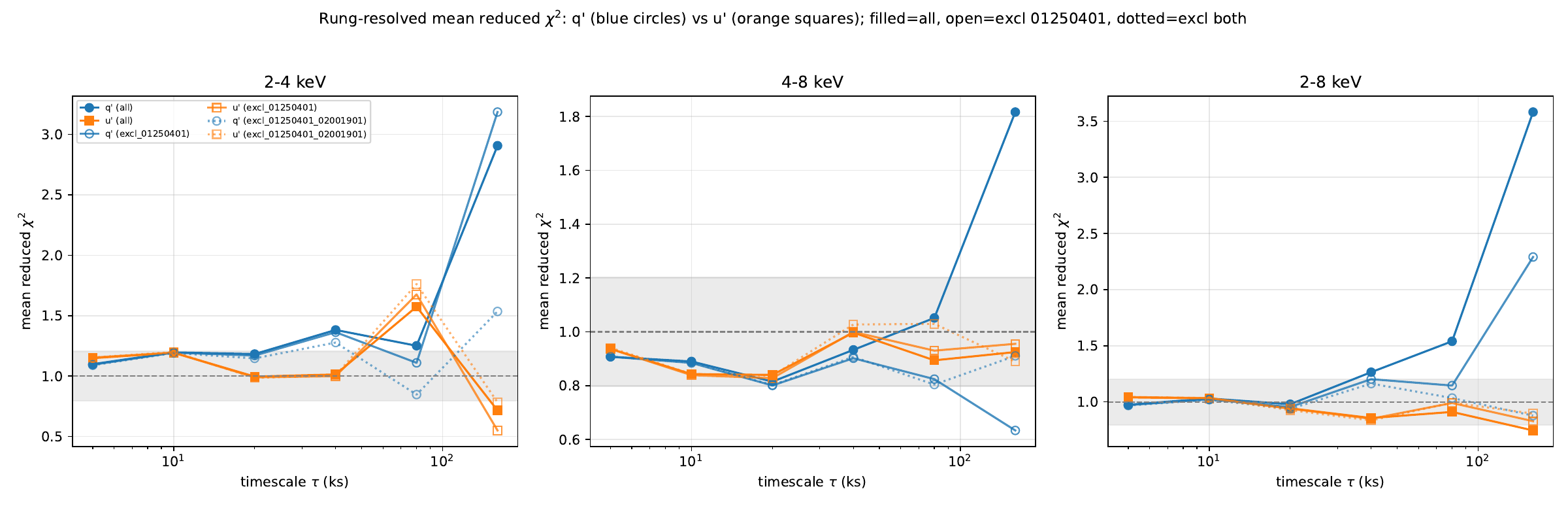}
 \caption{Mean reduced $\chisq$ of the constant fits of $q'$ (blue circles) and $u'$ (orange squares) as a function of the time binning and the energy band for all fits (filled), without the 4U~1630$-$47 observation 01250401 (open) and without both known variable observations (dotted); the grey band ($\pm0.20$) is the expected $1\sigma$ scatter of the reduced $\chisq$ of a single 5\,ks cell, $\sqrt{2/{\rm dof}}$ for the typical dof$\,\approx\,$48; the actual scatter of the six plotted means is 0.09 at 5\,ks and grows to $\sim$1 at 160\,ks, where only three cells contribute.}
\label{f:ecal}
\end{figure*}
As mentioned above, the pipeline attempted calibrating (scaling) the errors to deliver 
mean reduced $\chi^2$-values consistent with the expected value of 1. Figure \ref{f:ecal} shows the mean values before the scaling of the error bars for different energy bands and time binnings. 
Whereas the 2--8\,keV results have the 
right dispersion using the nominal error bars, the 
2--4\,keV $\chi^2$-values exceed 1 and the 
4--8\,keV $\chi^2$-values are below 1. 
The mean $\chi^2$-value correlates with neither the 
background rate nor the source flux, excluding the possibility that the deviations from the expected values can be explained with background or source rate dependent systematics.

\begin{figure}
\centering
\includegraphics[width=0.99\linewidth]{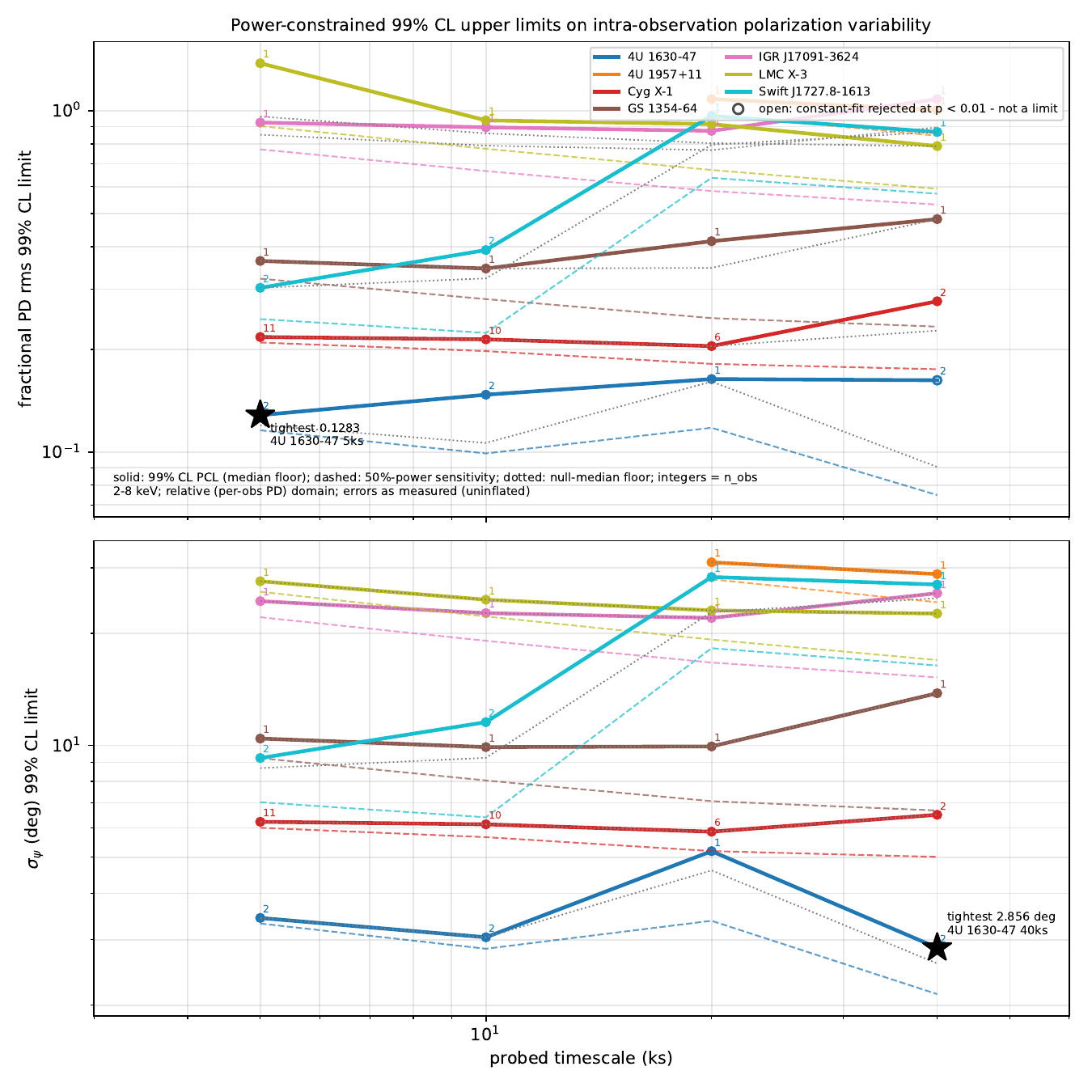}
\caption{Top: the power-constrained 99\% confidence upper limit on the RMS of the fractional variation of the polarization degree within an observation from bin to bin, versus the time scale probed, one thick line per source (lines to guide the eye; the small integers are the number of observations entering each point), with the 50\%-power sensitivity (dashed) and the null-median floor (dotted). Bottom: same for the bin-to-bin RMS wander of the polarization angle in degrees. Stars mark the tightest cell of each row. The open symbol (4U~1630$-$47 at 40~ks, top row; joint $\chisq/{\rm dof} = 51.0/24, p = 9\times10^{-4}$) marks the one cell in which the constant-fit hypothesis is rejected at $p<0.01$ 
and is not a limit.}
\label{fig:fig1}
\end{figure}

The second main result of the analysis is the set of limits on the RMS amplitude of the bin-to-bin PD and PA variations. The results for the strongly polarized sources (Fig.\,\ref{fig:fig1}) show that the PDs vary with an RMS of less than 12.8\% (4U~1630$-$47) to $\sim$100\% (LMC X-3 and IGR~J17091$-$3624) of their value. The PA wander is limited to RMSs smaller than 2.9$^{\circ}$ (4U~1630$-$47) to 30$^{\circ}$ (4U~1957$+$11). The sensitivity for detecting a certain RMS wander improves slowly with increasing time bin length (as the measurements become more precise but the number of bins shrinks); for some sources, e.g., 4U~1630$-$47 and Cyg X-1, the limits increase with time bin length as the observations start to pick up true source variability. For the sources with weak polarization detections, the power-constrained 99\% CL upper limits on the per-component RMS of $q$ and $u$ are $\sim$1\% (GX~339$-$4, Swift~J151857.0$-$572147) and 2.4\% (LMC X-1) at 5~ks (these limits are not shown in this paper).

\begin{figure*}
\centering
\includegraphics[width=\linewidth]{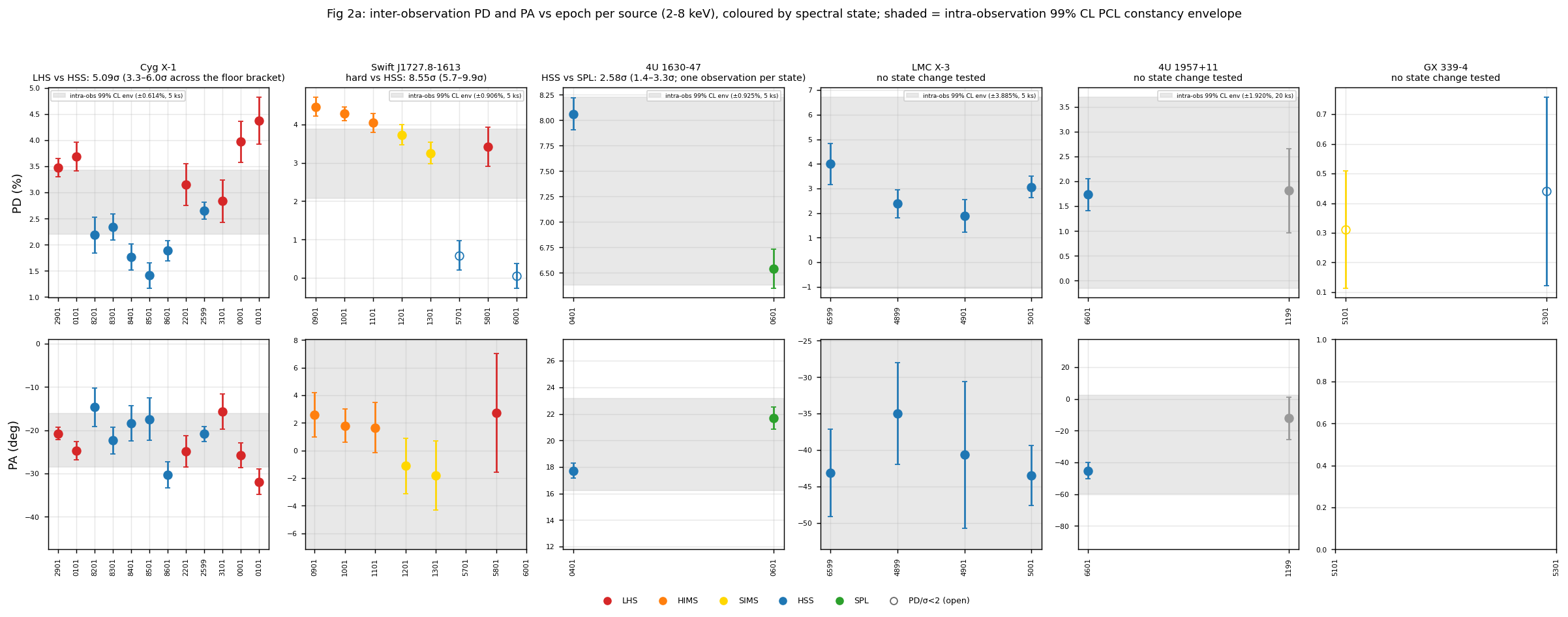}
\caption{Whole-observation polarization degree (top) and angle (bottom) of the sources with at least two observations, in time order, colored by spectral state; the panel titles give the significance of the difference between state groups after a detector-systematic error of 0.0039 in normalized Stokes has been added in quadrature to every observation (the floor, from the largest per-detector offset over 24 observations), followed by the same significance for the floor at the 16th and 84th percentiles of its bootstrap distribution, 0.0030 and 0.0066.
Shaded bands: the intra-observation 99\% confidence envelope at the shortest quotable time scale. The x-axis labels are the last four digits of the obs-IDs. Open circles mark observations whose PD is consistent with zero, for which the angle is undefined. Spectral states follow \citet{2026MNRAS.545f1933M}; the 2025 observation of 4U~1957$+$11 (04251199) postdates that classification and is shown in grey.}
\label{fig:fig2a}
\end{figure*}
In the fifth follow-up round, we asked the analysis agent to characterize the inter-observation PD and PA variability. The pipeline decided to characterize the variability as ``emission state dependent'', as the observations that disagreed with the mean were those in a different spectral state (spectral states taken from \citealt{2026MNRAS.545f1933M}), with PD tracking hardness. Fig.\,\ref{fig:fig2a} shows the result. 
Within a state, the observations are consistent with constant polarization, with 99\% CL limits 
on the excess RMS tighter than the between-state change, e.g.\ for Cyg X-1, the PD RMS is $<$0.30 (hard state) and $<$0.34 (soft state) of the PD and the PA RMS is $<8.7^{\circ}$ and $<9.6^{\circ}$; for Swift~J1727, the PD RMS is $<$0.15 of the PD and the PA RMS is $<4.3^{\circ}$ (hard states). 
Between states, the $(q, u)$ values differ at 5.1$\sigma$ (Cyg X-1) and 8.5$\sigma$ (Swift~J1727).
The statistical significances include the systematic errors between detector units which were added 
to the statistical errors in quadrature.

The run produced proposals for both knowledge databases. The recipe-learning agent proposed 14 analysis recipes over the 14 runs, among them the rotation of the Stokes parameters into the mean-vector frame, the per-band and per-time-scale check of the reduced $\chi^2$, the power-constrained limit with a null-median floor, the block-diagonal joint statistic for several observations of one source, the coverage test as a gate for quoting a limit, and the locking of headline numbers with a fixed random seed (to ensure reproducibility). The check-learning agent proposed 26 review checks. Although we accept most of them, three of the recipe proposals encode the wrong lesson and will not be merged: a ``limit'' recipe that is a detection sensitivity, a ``robustness test'' that drops the observations with the lowest PD (which removes the state change it is supposed to test), and two recipes about editing JSON files that belong in the orchestrator rather than in the knowledge base. 
Writing this paper produced another 11 recipes and 6 checks: compute every $\chi^2$, $p$-value and limit with the errors as measured and with inflated errors side by side; use the $q'$/$u'$ asymmetry to tell an error problem from real variability before calling it either; never sum $\chi^2$ over the rungs of a time-scale ladder, which re-bin the same data; divide counts by the exposure of exactly the data that were counted (all detector units, their good-time intervals, the true window edges); scan every time series for a slope before quoting a jitter limit; detrend and use a block-shuffled null before claiming a correlation between two slowly varying series; plot $-\log_{10}p$ of all tests against the null expectation with the trials thresholds, and rank the survivors with a stated sort; and give every surviving cell a light curve with the count rate and hardness next to $q'$ and $u'$. 
\section{Discussion and Outlook}
\label{s:discussion}
In this paper we present an AI pipeline for the analysis of the data from X-ray satellites. We find that the agentic system can analyze the data, and that the  review of the methods used and the results by the AI system itself and the scientist can be used to vet the results. 
The scientists and the AI agents can fruitfully collaborate in several stages of the research: the initial formulation and refinement of the analysis project, the audit of the methods and results, the fine tuning of the pipeline, and the interpretation of the results as well as the formulation of next steps. The agentic system is well suited to reduce the data, scrutinize it for consistency, and for applying statistical tests. 
Advisors of junior researchers are well aware that they sometimes discover
misunderstandings where the junior researcher did not apply a state-of-the-art method in a certain analysis step. The same can certainly happen when
using an agentic system to help with the analysis. As reading all the scripts that a junior researcher or an agentic system produces is not realistic, 
we believe one has to fall back to time tested methods to make sure that the analysis results are solid: include comparisons of some results with published results, redo several aspects of the analysis in several ways and see if the results behave as expected, and independently rederive some results and compare them with the results from the analysis pipeline. 

We applied this advice to the results shown in this paper: an independent provenance audit of Figs.~\ref{f:r36}--\ref{fig:fig2a}, performed with a Claude agent session and archived with the code, traced each figure to the level-2 event files, re-derived the quoted numbers from the result tables, and checked the transcription of the literature values against the source paper. All figures and numbers held up with the exception of several inaccuracies where the result descriptions deviated slightly from what the scripts actually did, in terms of the background-subtraction rule used, 
the treatment of observations with a poorly defined polarization angle, the weighted versus photon count rates, and a caption quoting an error-inflated $\chisq$. It also found that the significances in the panel titles of Fig.~\ref{fig:fig2a} were string literals in the plotting script, and that two result tables had been overwritten in place by follow-up scripts. 

We show exemplary results from the pipeline: a couple of analysis projects reproduce published results. We find good agreement with the literature - except for some of the legacy data sets where the addition of event weights now produces $\sim$15\% lower PDs. One larger analysis project scrutinized long observations for intra-observation PD and PA variability. 
The agent proposed to do this in terms of the normalized Stokes parameters in the frame in which $q'$ is aligned with the electric field polarization vector - an excellent method. 

The search revealed one smooth PD increase for 4U\,1630--47 in the soft state, and one PD wander for LMC X-1 and a remarkable absence of PA variations. From an astrophysical viewpoint the lack of PA variations 
is not too surprising: the accretion disk of a 21.2 solar mass black hole like Cyg X-1 evolves on a dynamical time scale of a few times the light-crossing time $\Delta \tau\,=\,GM/c^3 \approx 100\,\mu$s. The 5\,ks duration of the shortest {\it IXPE} time bins used in the analysis thus exceeds the dynamical time scale by $\sim 5\times 10^7$. 
The {\it IXPE} measurements are thus sensitive to a secular evolution of the accretion flow configuration but not to fluctuations.
Remarkably, the PD changes significantly 
between states, but not the PAs.
If a major fraction of X-rays in the soft state 
come indeed from close to the innermost stable circular orbit (ISCO), one would expect a clockwise (counterclockwise) rotation of the PA away from the black hole's and accretion disk angular momentum axis when seeing the accretion disk from above (below) \citep{2009ApJ...701.1175S,2012ApJ...754..133K}.
A state-dependent change of the PA has not yet been observed for any BHXRB.  

The analysis for the variability project took 14 pipeline runs to converge, and the scientist and an independent agent scrutinizing the results revealed several inaccuracies. We believe that additional analysis and interpretation rules can help to lead to faster convergence and more robust results. 
Furthermore, the pipeline tended to delve too deep into statistical minutia, producing a large number of results (230 figures, 300 tables) that make it difficult to validate the results. Future work will aim at enabling the pipeline to identify major results and to focus the resources on establishing these major results rather than to follow up on too many open threads.  

Agents based on large language models are rapidly becoming increasingly powerful and the Claude agent 
wrote $>$90\% of the pipeline code. In the near future, commercially available agents will be able 
to do analyses. The scientists and instrument teams will still need to provide expert knowledge 
on how to analyze and interpret the data. 
The pipeline discussed in this paper leverages the following methods: \\[0.2ex]
(1) We supplied knowledge about how to analyze the X-ray data making use of the brittle analysis tools provided by the instrument teams.\\[0.2ex]
(2) Earlier attempts to use agentic AI for the analysis of astronomical data reported problems with the hallucination or simulation of data and the difficulties to avoid this behavior by modifying the LLM prompts \citep{2024arXiv241200431L,2025arXiv251026887V}. We enforced the integrity of the analysis structurally: 
(i) the analysis plans must be based on the data runs from a machine readable data inventory, avoiding analysis plans not grounded in real data. (ii) An independent critic vets the analysis plan before implementation. (iii) After the analysis proceeded, we use an audit agent that scrutinizes the analysis scripts and analysis products to make sure that only actual data are used in the analysis and that no analysis shortcuts are taken which render the results moot. (iv) We do not allow the agent to read the audit code (no introspection) nor to see the results of the audit. (v) We do not allow the agent to change its own code. The five-pronged approach has proven to be highly successful. \\[0.2ex]
(3) We designed the pipeline so that the results are reproducible. This was accomplished by connecting the results to the methods and the data that were used to produce them. All results come from {\tt Python} scripts running on a local computer, which are stored alongside the results. \\[0.2ex]

The main lessons learned can be summarized as follows:
\begin{enumerate}
\item Coming up with new analysis project ideas is difficult. Upon reviewing the pipeline's projects, we find that the majority of them are not worth the LLM costs nor the scientist time to review them. 
\item Overall, agentic research may not cut down on the time it takes to carry through a single analysis project of a specific type. However, once the pipeline, with the help of scientists, has solved the obstacles for one analysis, an agentic pipeline makes it easy to perform similar analyses for similar objects and/or similar missions.       
\item Research requires an iterative approach as the outcome of an analysis project is rarely known before scrutinizing the data. For the time being, it is advantageous to keep scientists involved in these iterations to open up or limit the analysis space. 
\item The agentic system tended to embark on increasingly complicated statistical analyses. Future work should explore how to make agents evaluate and refocus or widen the scope of an analysis.  
\item We limited the analysis space to the statistical analysis of data without doing astrophysical modeling. It will be interesting to evolve the analysis pipeline to perform astrophysical modeling as well.           
\end{enumerate}

We gained several practical insights:
\begin{enumerate}
\item Using the online help of the analysis tools avoids version mismatches between the documentation and the tools that are actually available.
\item The agents are programmed to prefer delivering results over a report of a failure. Prompting the agents to prioritize null-results over compromised results helps.
\item We encourage the use of the analysis pipeline from a separate user account with limited access rights 
to protect sensitive information from the analysis agents.
\end{enumerate}
We will maintain the code base and the analysis and interpretation knowledge databases in a public git archive. In future, mission teams who provide the data and the analysis software might want to develop and maintain ``cheat sheets'' that explain to out-of-the-box AI agents how to analyze the data.
We will use a small sample of analysis projects to benchmark the performance of future iterations of the analysis pipeline. In future, mission teams 
may provide analysis project descriptions in a similar format that scientists can use to benchmark the costs and performance of their analysis pipelines.

We mentioned in the introduction Giacconi's historical move to make the data of the {\it Einstein} observatory available to the astronomical community. Following the quote from the introduction, the article continues ``We have therefore agreed to relinquish our exclusive rights to the data during the first year and to make the facility available
to all qualified observers at the end of the activation period (2 months after launch).'' Little did Giacconi and coworkers anticipate that ``qualified observers'' might someday include AI agents.
\section*{Acknowledgments}
The authors acknowledge support from the Physics Department at Washington University in St. Louis in the framework of the Hufferd Seed Fund program, as well as NASA support from the grant 80NSSC24K0205.
\section*{Use of AI}
The authors acknowledge the use of Claude (primary engine) and Gemini (secondary engine) for the development of the code of the analysis project.
We furthermore discussed the paper and the analysis chain with Claude and Gemini and implemented the feedback from these two models. 
The authors used Claude to help summarize the results of the pipeline in Section \ref{s:res}.
\section*{Data Availability}
The pipeline code and the data bases are publicly available under the name Agentic X-ray Analysis (AXA) at \url{https://gitlab.com/krawcz/AXA}.

\bibliographystyle{rasti}
\bibliography{ai_paper}

\label{lastpage}
\end{document}